\documentclass{sig-alternate}
\usepackage{hyperref}
\usepackage{enumitem}
\usepackage{cleveref}
\usepackage{algpseudocode}
\usepackage{algorithm}
\usepackage{balance}

\newtheorem{theorem}{Theorem}[section]
\newtheorem{lemma}[theorem]{Lemma}
\newtheorem{defin}{Definition}[section]
\newtheorem{proposition}[theorem]{Proposition}

\newlist{Axiom}{enumerate}{10}
\setlist[Axiom]{label=\textbf{R\arabic*},ref=(Rule \arabic*)}

\crefformat{section}{#2(\S#1)#3}
\crefformat{defin}{#2(Def.~#1)#3}
\crefformat{figure}{#2Fig.~#1#3}
\crefformat{theorem}{#2Theorem.~#1#3}

\usepackage[normal,bf]{caption}

\toappear{\em This is the extended version of the paper that appears in PODS 2013, New York, NY, USA}

\begin{document}
\clubpenalty=10000 
\widowpenalty = 10000

\renewcommand{\captionfont}{\bf \small}

\title{Learning and Verifying Quantified Boolean Queries by Example}
\numberofauthors{1}
\author{
\alignauthor Azza Abouzied$^*$, Dana Angluin$^*$, Christos Papadimitriou$^{**}$, \\ Joseph M. Hellerstein$^{**}$, Avi Silberschatz$^*$ \\
	   \affaddr{$^*$Yale University, $^{**}$ University of California, Berkeley}
       \email{azza@cs.yale.edu, angluin@cs.yale.edu, christos@cs.berkeley.edu, hellerstein@cs.berkeley.edu, avi@cs.yale.edu}
}

\maketitle
\begin{abstract}
To help a user specify and verify quantified queries --- a class of database queries known to be very 
challenging for all but the most expert users --- one can question the user on whether certain data objects are
\emph{answers} or \emph{non-answers} to her intended query. In this paper, we analyze the number of questions 
needed to learn or verify \emph{qhorn} queries, a special class of Boolean quantified queries whose underlying form is \emph{conjunctions of quantified Horn expressions}. We provide optimal \emph{polynomial-question} and \emph{polynomial-time} learning and verification algorithms for two subclasses of the class qhorn with upper constant limits on a query's \emph{causal density}.
\end{abstract}



\section{Introduction}

It's a lovely morning, and you want to buy a box of chocolates for your research group. You walk into a chocolate store and ask for ``a box with dark chocolates --- some sugar-free with nuts or filling''.  However, your server is a pedantic logician who expects first-order logic statements. In response to your informal query he places in front of you a hundred boxes!  Despite your frustration, you are intrigued: you open the first box only to find one dark, sugar-free chocolate with nuts and many other varieties of white chocolates that you didn't order.  You push it aside, indicating your disapproval, and proceed to the second.  Inside, you are wondering:  Is there hope that I can communicate to this person my needs through a sequence of such interactions?

Everyday, we request things from each other using informal and incomplete query specifications. Our casual interactions facilitate such under-specified requests because we have developed questioning skills that help us clarify such requests. A typical interlocutor might ask you about corner cases, such as the presence of white chocolates in the box, to get to a precise query specification by example. As requesters, we prefer to begin with an outline of our query --- the key properties of the chocolates --- and then make our query precise using a few examples. As responders, we can build a precise query from the query outline and a few positive or negative examples --- acceptable or unacceptable chocolate boxes.

Typical database query interfaces behave like our logician. SQL interfaces, for example, force us to formulate precise quantified queries from the get go. Users find quantified query specification extremely challenging~\cite{dataplay, reisner}. Such queries evaluate propositions over \emph{sets of tuples} rather than individual tuples, to determine whether a set as a whole satisfies the query. Inherent in these queries are (i) the grouping of tuples into sets, and (ii) the binding of query expressions with either existential or universal quantifiers. Existential quantifiers ensure that some tuple in the set satisfies the expression, while universal quantifiers ensure that all tuples in the set satisfy the expression.

To simplify the specification of quantified queries, we built DataPlay~\cite{dataplay}. DataPlay tries to mimic casual human interactions: users first specify the simple propositions of a query. DataPlay then generates a simple quantified query that contains all the propositions. Since this query may be incorrect, users can label query results as \emph{answers} or \emph{non-answers} to their intended query. DataPlay uses this feedback on example tuple-sets to fix the incorrect query. Our evaluation of DataPlay shows that users prefer example-driven query specification techniques for specifying complex quantified queries~\cite{dataplay}. Motivated by these findings, we set out to answer the question: \emph{How far can we push the example-driven query specification paradigm?} This paper studies the theoretical limits of using examples to \emph{learn} and to \emph{verify} a special sub-class of quantified queries, which we call \emph{qhorn}, in the hope of eventually making query interfaces more human-like.

\subsection{Our contributions}
We formalize a query learning model where users specify propositions that form the building blocks of a Boolean quantified query. A learning algorithm then asks the users \emph{membership questions}: each question is an example data object, which the user classifies as either an \emph{answer} or a \emph{non-answer}. After a few questions, the learning algorithm terminates with the unique query that satisfies the user's responses to the membership questions.  The key challenge we address in this paper is how to design a learning algorithm that runs in polynomial time, asks as few questions as possible and exactly identifies the intended query.

We prove the following:
\begin{enumerate}[leftmargin=0.5cm]
\item Learning quantified Boolean queries is intractable: A doubly exponential number of questions is required \cref{sec:intractable}. Within a special class of quantified Boolean queries known as \emph{qhorn} \cref{sec:qhorn}, we prove two subclasses are exactly and efficiently learnable: \emph{qhorn-1} \cref{sec:qhorn1} and its superset \emph{role-preserving qhorn} \cref{sec:rpqhorn} with constant limits on \emph{causal density} \cref{def:cd}. 
\item We design an optimal algorithm to learn qhorn-1 queries using $O(n \lg n)$ questions where $n$ is the number of propositions in a query \cref{sec:qhorn1learn}.
\item We design an efficient algorithm to learn role-preserving qhorn queries using $O(kn \lg n + n^{\theta + 1})$ questions where $k$ is \emph{query size} \cref{def:size}, and $\theta$ is causal density \cref{sec:rpqhornlearn}.
\end{enumerate}
We also formalize a query verification model where the user specifies an entire query within the role-preserving qhorn query class. A verification algorithm then asks the user a set of membership questions known as the \emph{verification set}. Each query has a unique verification set. The verification algorithm classifies some questions in the set as answers and others as non-answers. The query is incorrect if the user disagrees with any of the query's classification of questions in the verification set. 

We design a verification algorithm that asks $O(k)$ membership questions \cref{sec:verify}.

\section{Preliminaries}
\label{sec:prelim}
\label{sec:intractable}

Before we describe our query learning and verification algorithms, we first describe our data model --- \emph{nested relations} --- and the qhorn query class.

\begin{defin} Given the sets $D_1, D_2, ..., D_m$, $\mathcal{R}$ is a \textbf{relation} on these $m$ sets if it is a set of $m$-tuples $(d_1, d_2, ..., d_m)$ such that $d_i \in D_i$ for $i = 1, ..., m$. $D_1, ... , D_m$ are the the domains of $\mathcal{R}$. 
\end{defin}

\begin{defin} A \textbf{nested relation} $\mathcal{R}$ has at least one domain $D_i$ that is a set of subsets (powerset) of another relation $\mathcal{R}_i$. This $\mathcal{R}_i$ is said to be an embedded relation of $\mathcal{R}$.
\end{defin}

\begin{defin} A relation $\mathcal{R}$ is a \textbf{flat relation} if all its domains $D_1, ..., D_m$ are not powersets of another relation.\end{defin}

For example, a flat relation of chocolates can have the following schema:
\begin{equation*}
\begin{array}{l}
\texttt{Chocolate(isDark, hasFilling, isSugarFree,} \\
\texttt{hasNuts, origin)}
\end{array}
\end{equation*}

A nested relation of boxes of chocolates can have the following schema:
\begin{equation*}
\begin{array}{l}
\texttt{Box(name, Chocolate(isDark, hasFilling,}\\
\texttt{isSugarFree, hasNuts, origin))}
\end{array}
\end{equation*}
In this paper, we analyze queries over a nested relation with single-level nesting, i.e. the embedded relation is flat. The  \texttt{Box} relation satisfies single-level nesting as the \texttt{Chocolate} relation embedded in it is flat. To avoid confusion, we refer to elements of the nested relation as \emph{objects} and elements of the embedded flat relation as \emph{tuples}. So the boxes are objects and the individual chocolates are tuples.

\begin{defin}
A \textbf{Boolean query} maps objects into either \emph{answers} or \emph{non-answers}.
\end{defin}

The atoms of a query are Boolean propositions such as:
\begin{equation*}
\begin{array}{l}
p_1: c.\texttt{isDark}, \ p_2: c.\texttt{hasFilling},\\
p_3:$ $c.\texttt{origin = Madagascar}
\end{array}
\end{equation*}

A complete query statement assigns quantifiers to expressions on propositions over attributes of the embedded relation. For example: 
\begin{equation}
\label{eqn:choco}
\begin{array}{l}
\forall c \in \texttt{Box.Chocolates}\  (p_1) \ \wedge \\
\exists c \in \texttt{Box.Chocolates}\  (p_2 \wedge p_3)
\end{array}
\end{equation}

A box of chocolates is an answer to this query if every chocolate in the box is dark and there is at least one chocolate in the box that has filling and comes from Madagascar. 

\begin{figure}[htbp]
   \centering
   \includegraphics[width=1\linewidth]{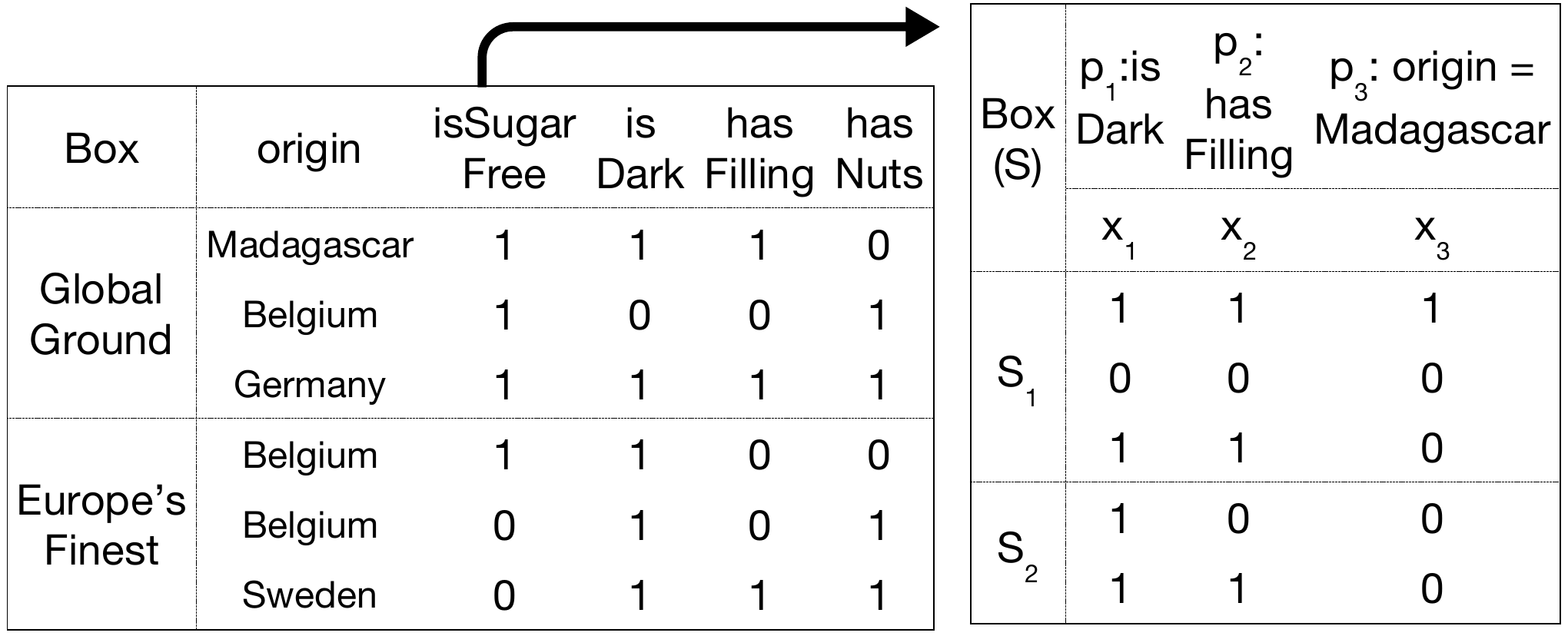}
   \caption{Transforming data from its domain into a Boolean domain.}
   \label{fig:transformation}
\end{figure}

Given a collection of propositions, we can construct an abstract Boolean representation for the tuples of the nested relation. For example, given propositions $p_1, p_2, p_3$, we can transform the chocolates from the data domain to the Boolean domain as seen in Figure \ref{fig:transformation}.

Thus, each proposition $p_i$ is replaced with a Boolean variable $x_i$. We rewrite the Boolean query (\ref{eqn:choco}) as follows:
\begin{equation*}
\begin{array}{l}
\label{choco}
\forall t \in S\  (x_1) \ \wedge \\
\exists t \in S\  (x_2 \wedge x_3)
\end{array}
\end{equation*}
where $S$ is the set of Boolean tuples for an object. This Boolean representation allows us to create learning and verification algorithms independent of the data domain or of the actual propositions that the user writes. 

To support this Boolean representation of tuples, however, we assume that (i) it is relatively efficient to construct an actual data tuple from a Boolean tuple and that (ii) the true/false assignment to one proposition does not interfere with the true/false assignments to other propositions. The propositions $p_m: c.\texttt{origin = Madagascar}$ and $p_b: c.\texttt{origin = Belgium}$ interfere with each other as a chocolate cannot be both from Madagascar and Belgium: $p_m \rightarrow \neg p_b$ and $p_b \rightarrow \neg p_m$. 

With three propositions, we can construct $2^3$ possible Boolean tuples, corresponding to the $2^3$ possible true or false assignments to the individual propositions, i.e. we can construct 8 different chocolate classes. With $n$ propositions, we can construct $2^n$ Boolean tuples.

There are $2^{2^n}$ possible sets of Boolean tuples or unique objects. With our three chocolate propositions, we can construct 256 boxes of distinct mixes of the 8 chocolate classes. Since a Boolean query maps each possible object into an answer or a non-answer, it follows that there are $2^{2^{2^n}}$ distinguishable Boolean queries (for $n=3$, about $10^{77}$). If our goal is to learn \emph{any} query from $n$ simple propositions by asking users to label objects as answers or non-answers, i.e. asking {\em membership questions}, then we would have to distinguish between $2^{2^{2^n}}$ queries using $\Omega(\lg(2^{2^{2^n}}))$ or $2^{2^n}$ questions. 

Since this ambitious goal of learning any query with few membership questions is doomed to fail, we have to constrain the query space. We study the learnability of a special space of queries, which we refer to as \emph{qhorn}.

\subsection{Qhorn}
\label{sec:qhorn}
Qhorn has the following properties:

\begin{enumerate}[leftmargin=0.5cm]
\item It supports if-then query semantics via quantified \emph{Horn} expressions: $\forall t\in S \  (x_1\wedge x_2 \rightarrow x_3)$. A Horn expression has a conjunction of body variables that imply a single head variable. The degenerate \emph{headless} Horn expression is simply a quantified conjunction of body variables ($\exists t\in S (x_1 \wedge x_2)$) and the degenerate \emph{bodyless} Horn expression is simply a single quantified variable ($\forall t\in S (\mathbf{T} \rightarrow x_1) \equiv \forall t\in S (x_1)$). 
\item It requires at least one positive instance for each Horn expression via a \emph{guarantee clause}. Thus, we add the existential clause $\exists t \in S \ (x_1\wedge x_2\wedge x_3)$ to the expression $\forall t\in S \  (x_1\wedge x_2 \rightarrow x_3)$ to get a complete query.  Note that the expression $\exists t\in S \  (x_1\wedge x_2 \rightarrow x_3)$ is implied by its guarantee clause $\exists t\in S \  (x_1\wedge x_2 \wedge x_3)$. 

We justify the naturalness of guarantee clauses with the following example: consider a user looking for a box of only sugar-free chocolates. Without the guarantee clause, an empty box satisfies the user's query. While such a result is logical, we contend that most users would not consider the result as representative of sugar-free chocolate boxes. 

\item It represents queries in a normalized form: \emph{conjunctions of quantified (Horn) expressions}. 
\end{enumerate}

We use a shorthand notation for queries in qhorn. We drop the implicit `$t \in S$', the `$\wedge$' symbol and the guarantee clause. Thus, we write the query 
\begin{equation*}
\begin{array}{l}
\forall t\in S \  (x_1\wedge x_2 \rightarrow x_3) \  \wedge \  \exists t \in S \ (x_1\wedge x_2\wedge x_3) \wedge \\
\forall t \in S \  (x_4) \ \wedge \ \exists t \in S \ (x_4) \ \wedge \ \exists t \in S \ (x_5)  
\end{array}
\end{equation*}
as
$\forall x_1x_2 \rightarrow x_3 \ \forall x_4 \ \exists x_5$.

\subsubsection{Qhorn's Equivalence Rules}
\label{sec:axioms}
\begin{Axiom}[leftmargin=0.5cm]
\item \label{ax:exi-dominate} The query representation $\exists x_1x_2x_3 \ \exists x_1x_2  \ \exists x_2x_3$ is equivalent to $\exists x_1x_2x_3$. This is because if a set contains a tuple that satisfies $\exists x_1x_2x_3$, that tuple will also satisfy $\exists x_1x_2$ and  $\exists x_2x_3$. An existential conjunction over a set of variables \textbf{dominates} any conjunction over a subset of those variables. 

\item \label{ax:uni-dominate} The query representation $\forall x_1x_2x_3 \rightarrow h \ \forall x_1x_2 \rightarrow h  \ \forall x_1 \rightarrow h$ is equivalent to $\forall x_1 \rightarrow h \ \exists x_1x_2x_3 \rightarrow h$. This is because $h$ has to be true whenever $x_1$ is true regardless of the true/false assignment of $x_2,x_3$. Thus a universal Horn expression with body variables $B$ and head variable $h$ \textbf{dominates} any universal Horn expression with body variables $B'$ and head variable $h$ where $B' \supseteq B$. 

\item \label{ax:exi-extend} The query representation $\forall x_1 \rightarrow h \ \exists x_1x_3$ is equivalent to $\forall x_1 \rightarrow h \ \exists x_1x_3h$. Again, this equivalence is because $h$ has to be true whenever $x_1$ is true.
\end{Axiom}

\subsubsection{Learning with Membership Questions}
\label{sec:mq}

A \textbf{membership question} is simply an object along with its nested data tuples. The user responds to such a question by classifying the object as an answer or a non-answer for their intended query. 

Given a collection of $n$ propositions on the nested relation, the learning algorithm constructs a membership question in the Boolean domain: a set of Boolean tuples on $n$ Boolean variables $x_1,...,x_n$ --- a variable for each proposition. Such a set is transformed into an object in the data domain before presentation to the user.

For brevity, we describe a membership question in the Boolean domain only. As a notational shorthand, we use $1^n$ to denote a Boolean tuple where all variables are true. We use lowercase letters for variables and uppercase letters for sets of variables. 

The following definitions describe two structural properties of qhorn queries that influence its learnability:

\begin{defin}\label{def:size} \textbf{Query size}, $k$, is the number of expressions in the query. \end{defin}

\begin{defin}\label{def:cd} \textbf{Causal Density}, $\theta$, is the maximum number of distinct non-dominated universal Horn expressions for a given head variable $h$. \end{defin}

Conceptually, universal Horn expressions represent causation: whenever the body variables are true, the head variable has to be true. If a head variable has many universal Horn expressions, it has many causes for it to be true and thus has a high causal density. 

The following inequality between causal density, $\theta$ and query size $k$ holds:
$0 \leq \theta \leq k$. We would expect users' queries to be small in size $k = O(n)$ and to have low causal density $\theta$.

A query class is \emph{efficiently} learnable if (i) the number of membership questions that a learning algorithm asks the user is polynomial in the number of propositions $n$ and query size $k$ and (ii) the learning algorithm runs in time polynomial in $n$ and $k$. Question generation needs to be in polynomial time to ensure interactive performance. This requirement entails that the number of Boolean tuples per question is polynomial in $n$ and $k$.  A query class is \emph{exactly} learnable if we can learn the exact target query that satisfies the user's responses to the membership questions. 

Due to the following theorem, qhorn cannot be efficiently and exactly learned with a tractable number of questions (even when query size is polynomially bounded in the number of propositions ($k = n$) and causal density has an upper bound of one ($\theta = 1$)).

\begin{theorem}
\label{thm:intract}
Learning qhorn queries where variables can repeat $r \geq 2$ times requires $\Omega(2^n)$ questions.
\end{theorem}

\emph{Proof:}
Suppose we split our $n$ variables into two disjoint subsets: $X, Y$. Consider the query class: 
\begin{equation*}
\begin{array}{rcl}
\phi &=& \textrm{Uni}(X) \wedge \textrm{Alias}(Y)\\
\textrm{Uni}(X) &=& \forall x_1 \ \forall x_2 ... \forall x_{|X|}\\
\textrm{Alias}(Y) &=& \forall y_1 \rightarrow y_2 \ \forall y_2 \rightarrow y_3 ... \forall y_{|Y|} \rightarrow y_1
\end{array}
\end{equation*}

$\phi$ is simply the class of qHorn queries where some variables, $X$, are universally quantified and bodyless and the other variables $Y$ form an \emph{alias} i.e. all variables in $Y$ are either all true or all false. An example instance from this class of queries over variables $\{x_1, x_2, ..., x_6\}$ is: $\textrm{Uni}({\{x_1, x_3, x_5\})} \wedge \textrm{Alias}({\{x_2,x_4,x_6\})} = \forall x_1 \forall x_3 \forall x_5 \  \wedge \ \forall x_2 \rightarrow x_4 \forall x_4 \rightarrow x_6 \forall x_6 \rightarrow x_2$. Only two questions satisfy the example instance (i) a question with only the Boolean tuple $1^{6}$ and (ii) a question with tuples: $\{1^{6}, 101010\}$.

There are $2^{n}$ query instances in the class $\phi$. If we construct a membership question with only $1^{n}$ tuples then for all instances in $\phi$, the question is an answer and we cannot learn the target query. If we augment the question with one tuple where some variables are false then all such questions are non-answers unless exactly the false variables are the alias variables. Augmenting the question with two or more tuples will always be a non-answer even if one of the tuples has exactly and only the alias variables set to false. This is because the other tuples either have some universally-quantified bodyless variables that are false or some alias variables that are true and some alias variables that are false. 

This leaves us with $2^n$ membership questions where each question satisfies exactly one target query. Consider an adversary who always responds `non-answer'. In the worst case we have to ask $2^n - 1$ questions as each question eliminates exactly one query from consideration as the target query. \qed

Qhorn's intractability does not mean that we cannot construct efficiently and exactly learnable qhorn subclasses. We describe two such sub-classes:
\subsubsection{Qhorn-1}
\label{sec:qhorn1}

Qhorn-1 defines certain syntactic restrictions on qhorn. Not counting guarantee clauses, if a query has $k$ distinct expressions ($1 \leq k \leq n$) and each expression $i$ has body variables $B_i$ and a head variable $h_i$, such that $B = B_1 \cup ... \cup B_k$ is the collection of all body variables and $H = \{h_1, ... h_m\}$ is the set of all head variables then the following restrictions hold in qhorn-1:
\begin{enumerate}[noitemsep]
\item $B_i \cap B_j = \emptyset \vee B_i = B_j \textrm{ if } i \neq j$
\item $h_i \neq h_j \textrm{ if } i \neq j$
\item $B \cap H = \emptyset$
\end{enumerate}

The first restriction ensures that different head variables can either share the exact same set of body variables or have disjoint bodies. The second restriction ensures that a head variable has only one body. Finally, the third restriction ensures that a head variable does not reappear as a body variable. Effectively, qhorn-1 has \emph{no variable repetition}: a variable can appear \textbf{once} either in a set of body variables or as a head variable. The following diagram labels the different components of a qhorn-1 query.

\begin{figure}[h]
   \centering
   \includegraphics[width=0.7\linewidth]{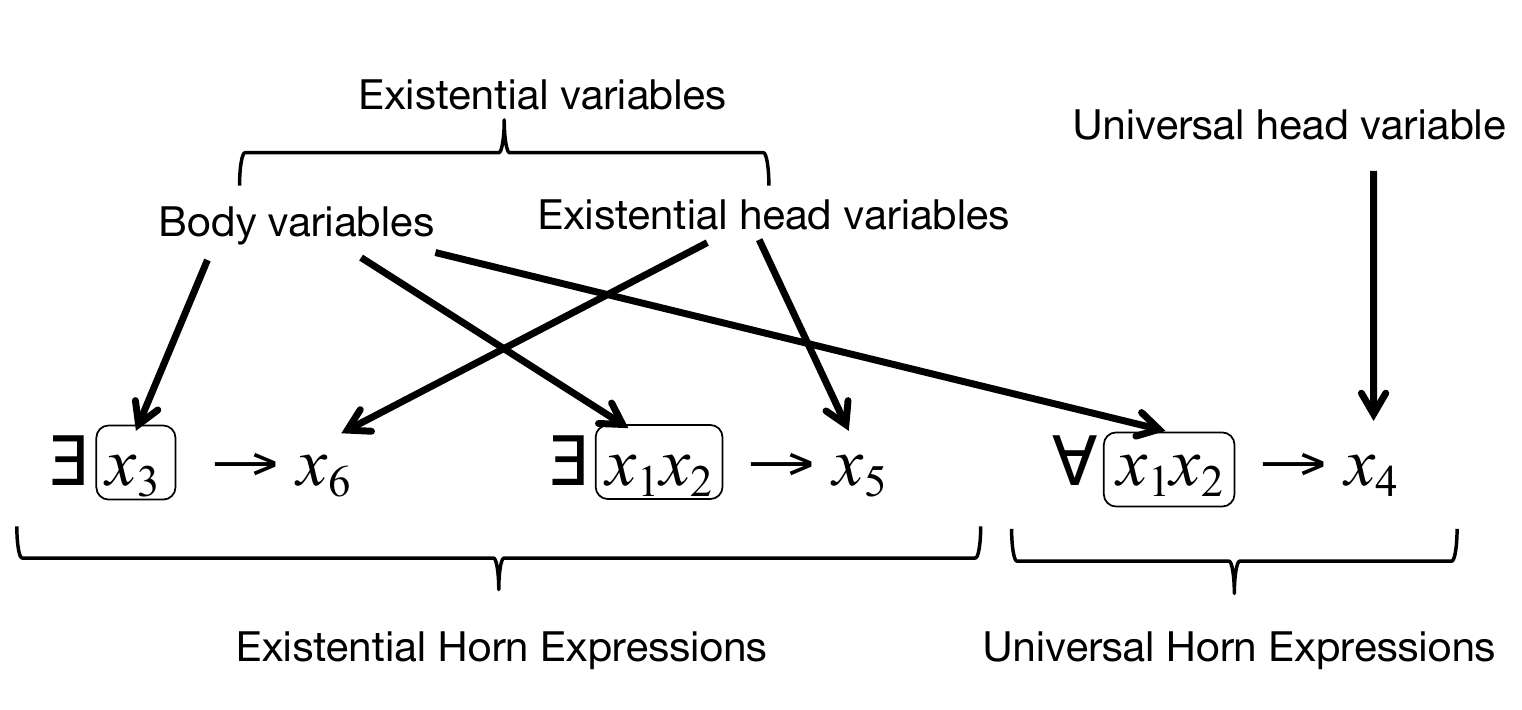}
   \caption{The different components of a qhorn-1 query.}
   \label{fig:qhorn1labels}
\end{figure}

Note that qhorn-1 queries have a maximum query size $k$ of $n$ and have a causal density $\theta$ of at most one. 
From an information-theoretic perspective, $\Omega(n \lg n)$ membership questions are required to learn a target query in qhorn-1.
This is because qhorn-1 has $2^\Theta({n \lg n})$ queries. We can think of all head variables that share the same set of body variables to be one part of a partition of the $n$ Boolean variables. If we can construct a unique qhorn-1 query from every partition of $n$ variables, then a lower bound on the number of queries is the Bell Number $B_n$, i.e. the number of ways we can partition a set into non-empty, non-overlapping subsets. One way to construct a unique query for every partition is as follows:

\begin{enumerate}
\item We universally quantify all variables that appear in a singleton part: $\forall x_i$
\item For all other parts, we pick any one variable as the head of an existentially quantified Horn expression with the remaining variables as body variables.
\end{enumerate}

For example, we construct the query $\forall x_1 \ \forall x_2$ $\exists x_3\rightarrow x_4$ $\exists x_5x_6 \rightarrow x_7$ from the partition $x_1 | x_2 | x_3x_4 | x_5x_6x_7$. Since $\ln(B_n) = \Theta(n \ln n)$, a lower bound estimate on the number of queries in qhorn-1 is $2^{n \lg n}$. 

Note that for each part, we can have either an existential or a universal quantifier and we can set a variable's \emph{role} as either a head or a body variable. Since, we can have most $n$ parts, an upper bound estimate on the number of queries is $2^n \times 2^n \times 2^{n \lg n}$. Thus the size of qhorn-1 is $2^{\Theta(n \lg n)}$.

\subsubsection{Role-preserving qhorn}
\label{sec:role}
\label{sec:rpqhorn}

In role-preserving qhorn queries, variables can repeat many times, but across universal Horn expressions head variables can only repeat as head variables and body variables can only repeat as body variables. For example, the following query is in role-preserving qhorn
\[\forall x_1x_4 \rightarrow x_5 \ \forall x_3x_4 \rightarrow x_5 \  \forall x_2x_4 \rightarrow x_6 \  \exists x_1x_2x_3  \ \exists x_1x_2x_5x_6 \]
while the following query is not in role-preserving qhorn
\[\forall x_1x_4 \rightarrow x_5 \  \forall x_2x_3x_5 \rightarrow x_6\]
because $x_5$ appears both as a head variable and a body variable in two universally quantified Horn expressions. Existential Horn expressions in role-preserving qhorn are rewritten as existential conjunctions and variables do not have \emph{roles} in these conjunctions. Thus, existential conjunctions can contain one or more head variables (e.g. $\exists x_1x_2x_5x_6$ in the first query). The following diagram labels the different components of a role-preserving qhorn query.
\begin{figure}[h]
   \centering
   \includegraphics[width=0.7\linewidth]{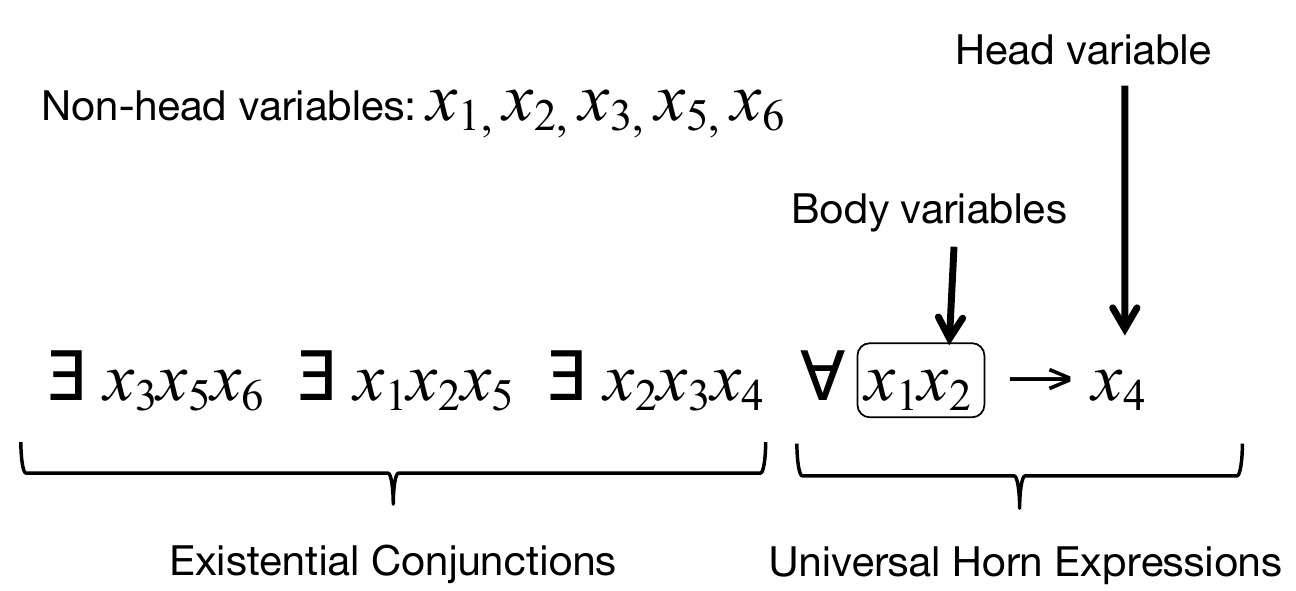}
   \caption{The different components of a role-preserving qhorn query.}
   \label{fig:rpqhornlabels}
\end{figure}

Both query size and causal density play a role in the behavior of learning and verification algorithms. Once we remove the syntactic restriction of variables appearing at most once, the size of a target query instance is no longer polynomially bounded in $n$. Thus, the complexity of learning and verification algorithms for role-preserving qhorn queries is parameterized by $k$, $\theta$ and $n$. We would expect user queries to have low causal densities and to be small in size. Provided that $\theta$ has a constant upper bound, then we can efficiently learn role-preserving queries.

\section{Query Learning}

\subsection{Learning qhorn-1}
\label{sec:qhorn1learn}

\begin{theorem}
$O(n \lg n)$ questions are sufficient to learn qhorn-1 queries in polynomial time.
\end{theorem}
\emph{Proof:} The learning algorithm breaks down query learning into a series of small tasks. First, it classifies all variables into either \emph{universal head} variables or \emph{existential} variables (\cref{fig:qhorn1labels} describes qhorn-1 terminology). Second, it learns the body variables (if any) for each universal head variable. Finally, it learns existential Horn expressions. We show that each task requires at most $O(n \lg n)$ membership questions (Section \ref{sec:qtypes}, Lemmas \ref{lem:univ-dep} and \ref{lem:matrix}), thus proving that the learning algorithm asks $O(n \lg n)$ questions.\qed

\subsubsection{Learning universal head variables}
\label{sec:qtypes}
The simplest learning task is to determine whether a variable is a universal head variable. Suppose we have three variables: $x_1,x_2,x_3$. To determine if $x_1$ is the head of a universal Horn expression, we ask the user if the set $\{111, 011\}$ is an answer. By setting the other variables ($x_2,x_3$) to be always true, we are setting all potential body variables of $x_1$ to true. We are also neutralizing the effect of other unknown head variables on the outcome of a membership question. If the set $\{111, 011\}$ is an answer, then we are sure that $x_1$ is not a universal head variable because it can exist with a false value as long as at least one tuple has a true value for it. If the set is a non-answer, then we learn that $x_1$ is a universal head variable. 

We need one question to determine whether a variable is a universal head variable and we need $O(n)$ time to generate each question --- the time to construct a set with two tuples of size $n$. Thus, we learn which variables are universal head variables, $U$, and which variables are existential variables, $E$, in polynomial time.

\subsubsection{Learning body variables of universal Horn expressions}
\label{sec:unibodies}

\begin{algorithm}[htb]                     
\small
\caption{\small Find bodies of universal head variable}          
\label{qhorn1uni}                           
\begin{algorithmic}                    
	\State $h$: The universal head variable
	\State $E$: The set of existential variables
	\State $\mathcal{B}$: is the set of bodies learned so far $(B_1, B_2, ...)$
	\State
	\State $b \leftarrow$ \textbf{Find}(UniversalDependence($h$, ?), Non-Answer, $B_1 \cup B_2 \cup ... \cup B_{|\mathcal{B}|}$)
	\If{$b \neq \emptyset$}
		\For{$B \in \mathcal{B}$}
			\If{$b \in B$}
				\State \Return $B$
			\EndIf
		\EndFor
	\EndIf
	\State $B \leftarrow$ \textbf{FindAll}(UniversalDependence($h$, ?), Non-Answer, $E$)
	\State \Return $B$
	\end{algorithmic}
\end{algorithm}

\begin{defin} Given a universal head variable $h$ and a subset of existential variables $V \subseteq E$, a \textbf{universal dependence question} on $h$ and $V$ is a membership question with two tuples: $1^n$ and a tuple where $h$ and $V$ are false and all other variables are true.
\end{defin}

If a universal dependence question on $h$ and $V$ is an answer, then we learn that a subset of $h$'s body variables is in $V$. This is because when the conjunction of body variables is not satisfied, the head variable can be false. We say that $h$ \emph{depends} on some variables in $V$. If the question is a non-answer, then we learn that $h$'s body variables are a subset of $E - V$; $h$ has no body variables in $V$ because in qhorn-1, $h$ can have at most one body.

The most straightforward way to learn the body variables, $B$, of one universal variable is with $O(|E|) = O(n)$ universal dependence questions: we serially test if $h$ depends on each variable $e \in E$. This means we use $O(n^2)$ questions to determine the body variables for all universal variables. We can do better.

\begin{algorithm}[htb]                     
\small
\caption{\small Find}          
\label{find}                           
\begin{algorithmic}                    
	\State $\mathbf{Q}$: The question to ask
	\State $r$: The response on which we eliminate a set of variables from further consideration
	\State $V$: The variables to apply binary search within
	\State
	\If{Ask($\mathbf{Q}$($D$)) = $r$}
		\State \Return $\emptyset$
	\Else
		\If{$|D| = 1$}
			\State \Return $D$
		\Else
			\State Split $D$ into $D_1$ (1$^{st}$ half) and $D_2$ (2$^{nd}$ half)
			\State $x \leftarrow$ Find($\mathbf{Q}$, $r$, $D_1$)
			\If{$x = \emptyset$}
				\State \Return Find($\mathbf{Q}$, $r$, $D_2$)
			\Else
				\State \Return $x$
			\EndIf
		\EndIf
	\EndIf
\end{algorithmic}
\end{algorithm}
\begin{algorithm}[htb]                      
\small
\caption{\small FindAll}          
\label{findall}                           
\begin{algorithmic}                    
	\State $\mathbf{Q}$: The question to ask
	\State $r$: The response on which we eliminate a set of variables from further consideration
	\State $V$: The variables to apply binary search within
	\State
	\If{Ask($\mathbf{Q}$($D$)) = $r$}
		\State \Return $\emptyset$
	\Else
		\If{$|D| = 1$}
			\State \Return $D$
		\Else
			\State Split $D$ into $D_1$ (1$^{st}$ half) and $D_2$ (2$^{nd}$ half)
			\State \Return FindAll($\mathbf{Q}$, $r$, $D_1$) $\cup$ FindAll($\mathbf{Q}$, $r$, $D_2$)
		\EndIf
	\EndIf
\end{algorithmic}
\end{algorithm}

We perform a binary search for $h$'s body variables in $E$. If $h$ has $B$ body variables, we ask $O(|B| \lg n)$ instead of $O(n)$ questions to determine $B$. Suppose we have four variables $x_1,x_2,x_3,x_4$ such that $x_1$ is a universal head variable and all other variables are existential variables. $x_2,x_3,x_4$ are potential body variables for $x_1$. If the set $\{1^n, 0^n\}$ is a non-answer then $x_1$ is independent of all other variables and it has no body. If the set is an answer, we divide and conquer the variables. We ask if $x_1$ universally depends on half the variables, $\{x_2,x_3\}$, with the set $\{1^n, 0001\}$. If the set is a non-answer then we eliminate half the variables, $\{x_2,x_3\}$, from further consideration as body variables. We know that a body variable has to exist in the remaining half and since, $x_4$ is the last remaining variable, we learn the expression $\forall x_4 \rightarrow x_1$. If the set $\{1^n, 0001\}$ is an answer, then we know at least one body variable exists in $\{x_2,x_3\}$ and we continue the search for body variables in $\{x_2,x_3\}$, making sure that we also search the other half $\{x_4\}$ for body variables.

\begin{lemma} \label{lem:univ-dep}$O(n \lg n)$ universal dependence questions are sufficient to learn the body variables of all universal head variables.\end{lemma}
\emph{Proof:} Suppose we partition all variables into $m$ non-overlapping parts of sizes $k_1,k_2,...,k_m$ such that $\sum_{i = 1}^{m}k_i = n$. Each part has at least one body variable and at least one universal head variable. Such a query class is in qhorn-1 as all body variables are disjoint across parts and head variables cannot reappear as head variables for other bodies or in the bodies of other head variables. 

Given a head variable $h_i$, we can determine its body variables $B_i$ using the binary search strategy above: we ask $O(|B_i| \lg n)$ questions (it takes $O(\lg n)$ questions to determine one body variable). For each additional head variable, $h_i'$, that shares $B_i$, we require at most $1\lg n$ questions: we only need to determine that $h_i'$ has one body variable in the set $B_i$. Thus to determine all variables and their roles in a part of size $k_i$ with $|B_i|$ body variables and $|H_i|$ head variables we need $O(|B_i|\lg n + |H_i|\times1\lg n) = O(k_i \lg n)$ questions. Since there are $m$ parts, we ask a total of $O(\sum_{i=1}^{m}k_i\lg n) = O(n \lg n)$ questions. \qed

Since universal dependence questions consist of two tuples we only need $O(n)$ time to generate each question. Thus, the overall running time of this subtask is in polynomial time.

\subsubsection{Learning existential Horn expressions}

\begin{algorithm}[htb]                      
\small
\caption{\small Learn existential Horn expressions}          
\label{exihorn}                           
\begin{algorithmic}                    
	\State $\mathcal{Q}$: The target qhorn-1 query
	\State $\mathcal{B}$: The set of bodies learned so far $(B_1, B_2, ...)$
	\State $E$: The set of all existential variables
	\State
	\For{$e \in \{E - (B_1 \cup B_2 \cup ... \cup B_{|\mathcal{B}|})\}$}
		\State $b \leftarrow $ \textbf{Find}(ExistentialIndependence($e$, ?), Answer, $B_1 \cup B_2 \cup ... \cup B_{|\mathcal{B}|}$)
		\If{$b \neq \emptyset$}
			\For{$B \in \mathcal{B}$}
				\If{$b \in B$}
					\State $\mathcal{Q} \leftarrow \mathcal{Q} \wedge \exists B \rightarrow e$
				\EndIf
			\EndFor
		\Else
			\State $D \leftarrow$ \textbf{FindAll}(ExistentialIndependence($e$, ?), Answer, $E$)
			\State $H \leftarrow$ GetHead($e, D$)
			\If{$H = \emptyset$}
				\State $\mathcal{Q} \leftarrow \mathcal{Q} \wedge \exists D \rightarrow e$
				\State $\mathcal{B} \leftarrow \mathcal{B} \cup \{D\}$
			\Else
				\State $h \leftarrow H[1]$
				\For{$d \in \{D - H\}$}
					\If{Ask(ExistentialIndependence($h$, $d$))}
						\State $H \leftarrow H \cup d$ 
					\EndIf
				\EndFor
				\State $B \leftarrow (D - H) \cup \{e\}$
				\State $\mathcal{B} \leftarrow \mathcal{B} \cup \{B\}$
				\For{$h \in H$}
					\State $\mathcal{Q} \leftarrow \mathcal{Q} \wedge \exists B \rightarrow h$
				\EndFor
			\EndIf
			\State $E \leftarrow E - D$
		\EndIf
	\EndFor
\end{algorithmic}
\end{algorithm}

After learning universal Horn expressions, we have established some non-overlapping distinct bodies and their universal head variables.  Each variable in the remaining set of existential variables, can either be (i) an existential head variable of one of the existing bodies or (ii) an existential head variable of a new body or (i) a body variable in the new body. We use \emph{existential independence} questions to differentiate between these cases.

\begin{defin} Given two disjoint subsets of existential variables $X \subset E, Y \subset E, X \cap Y = \emptyset$, an \textbf{existential independence question} is a membership question with two tuples: (i) a tuple where all variables $x \in X$ are false and all other variables are true and (ii) a tuple where all variables $y \in Y$ are false and all other variables are true.
\end{defin}

If an independence question between two existential variables $x$ and $y$ is an answer then either:
\begin{enumerate}[noitemsep,nolistsep]
\item $x$ and $y$ are existential head variables of the same body
\item or $x$ and $y$ are not in the same Horn expression.
\end{enumerate}
We say that $x$ and $y$ are \emph{independent} of each other. Two sets $X$ and $Y$ are independent of each other if all variables $x \in X$ are independent of all variables $y \in Y$. Conversely, if an independence question between $x$ and $y$ is a non-answer then either:
\begin{enumerate}[noitemsep,nolistsep]
\item $x$ and $y$ are body variables in the same body or
\item $y$ is an existential head variable and $x$ is in its body or
\item $x$ is an existential head variable and $y$ is in its body
\end{enumerate}
We say that $x$ and $y$ \emph{depend} on each other. If sets $X$ and $Y$ depend on each other then at least one variable $x \in X$ depends on one variable $y \in Y$.

Given an existential variable $e$, if we discover that $e$ depends on a body variable $b$ of a known set of body variables $B$, then we learn that $e$ is an existential head variable in the Horn expression: $\exists B \rightarrow e$. 

Otherwise, we find all existential variables $D$ that $e$ depends on. We can find all such variables with $O(|D| \lg n)$ existential independence questions using the binary search strategy of Section \ref{sec:unibodies}. 

Knowing that $D$ depends on $e$ only tell us that one of the following holds: (i) A subset $H$ of $D$ are existential head variables for the body of $e \cup (D - H)$ or (ii) $e$ is a head variable and $D$ is a body. To differentiate between the two possibilities we make use of the following rule: \emph{If two variables $x, y$ depend on $z$ but $x$ and $y$ are independent then $z$ is a body variable and $x, y$ are head variables}. If we find a pair of independent variables $h_1, h_2$ in $D$, we learn that $x$ must be a body variable. If we do not find a pair of independent variables in $D$ then we may assume that $x$ is an existential head variable and all variables in $D$ are body variables.

After finding head variables in $D$, we can determine the roles of the remaining variables in $D$ with $|D| = O(n)$ independence questions between $h_1$ and each variable $d \in D - h_1$.  If $h_1$ and $d$ are independent then $d$ is an existential head variable, otherwise $d$ is a body variable.

Our goal, therefore, is to locate a definitive existential head variable in $D$ by searching for an independent pair of variables. 

\begin{defin}
An \textbf{independence matrix question} on $D$ variables consists of $|D|$ tuples. For each variable $d \in D$, there is one tuple in the question where $d$ is false and all other variables are true.
\end{defin}
Suppose we have four variables $x_1,...,x_4$; $D = \{x_2,x_3,x_4\}$ and $D$ depends on $x_1$. $\{1{\bf011}, 1{\bf101}, 1{\bf110}\}$ is a matrix question on $D$. If such a question is an answer then there is \emph{at least a pair} of head variables in $D$: the question will always contain a pair of tuples that ensure that each head and the body is true. For example if $x_2, x_4$ are head variables then tuples $\{1{\bf011}, 1{\bf110}\}$ in the question satisfy the Horn expressions: $\exists x_1x_3 \rightarrow x_2, \exists x_1x_3 \rightarrow x_4$. If \emph{at most one} variable in $D$ is a head variable, then there is no tuple in the matrix question where all body variables are true and the head variable is true and the question is a non-answer. For example, if only $x_4$ is a head variable, then the tuple, $1111$ that satisfies the Horn expression $\exists x_1x_2x_3 \rightarrow x_4$ is absent from the question. 

\begin{lemma}
\label{lem:matrix}
Given an existential variable $x$ and its dependents $D$, we can find an existential head variable in $D$ with $O(|D| \lg |D|)$ independence matrix questions of $O(|D|)$ tuples each if at least two head variables exist in $D$.
\end{lemma}

\begin{algorithm}[htb]                      
\caption{\small Get Head}          
\label{alg1}                           
\small
\begin{algorithmic}                    
    \State $x$: an existential variable
    \State $D$: the dependents of $x$, $|D| \geq 1$
    \State $D_1 \leftarrow D, D_2 \leftarrow  \emptyset, D_3 \leftarrow  \emptyset$
    \While{$D_1 \neq \emptyset$}
    \State isAnswer $\leftarrow$ Ask(MatrixQuestion($x$, $D_1$))
    \If{isAnswer}
    	\If{$|D_1| = 2 \wedge D_2 = \emptyset$}
			\Return $D_1$
		\ElsIf{$|D_1| > 2 \wedge D_2 = \emptyset$}
			\State Split $D_1$ into $D_1$ (1$^{st}$ half) and $D_3$ (2$^{nd}$ half)
		\ElsIf{$|D_2| = 1$}
			\Return $D_2$
		\Else
			\State Split $D_2$ into $D_2$ (1$^{st}$ half) and $D_3$ (2$^{nd}$ half)
			\State $D_1 \leftarrow D_1 - D_3$
		\EndIf
    \Else
    	\If{$D_3 = \emptyset$}
			\Return $\emptyset$
		\ElsIf{$|D_3| = 1$}
			\Return $D_3$
		\Else
			\State Split $D_3$ into $D_2$ (1$^{st}$ half) and $D_3$ (2$^{nd}$ half)
			\State $D_1 \leftarrow D_1 \cup D_2$
    	\EndIf
    \EndIf
    \EndWhile
\end{algorithmic}
\end{algorithm}

\emph{Proof.} Consider the `GetHead' procedure in Alg.~\ref{alg1} that finds an existential head variable in the set $D$ of dependents of variable $x$. 
The central idea behind the `GetHead' procedure is if the user responds that a matrix question on $D_1$ ($D_1 \subseteq D$) is an answer, then a pair of head variables must exist in $D_1$ and we can eliminate the remaining variables $D - D_1$ from further consideration. Otherwise, we know that at most one head variable exists in $D_1$ and another exists in $D - D_1$ so we can eliminate $D_1$ from further consideration and focus on finding the head variable in $D - D_1$. 

Each membership question eliminates half the variables from further consideration as head variables. Thus, we require only $O(\lg |D|) = O(\lg n)$ questions to pinpoint one head variable. 

Then, we ask $O(|D|)$ questions to differentiate head from body variables in $D$. If we do not find head variables in $|D|$ then we may assume that $x$ is a head variable and all variables in $D$ are body variables. Once we learn one existential Horn expression, we process the remaining existential variables in $E$. If a variable depends on any one of the body variables, $B$, of a learned existential Horn expression, it is a head variable to all body variables in $B$. 

Suppose a query has $m$ distinct existential expressions with $k_1,...,k_m$ variables each, then $\sum_{i=1}^{m}k_i < n$.  The size of each set of dependent variables for each expression $i$ is $k_i - 1$. So the total number of questions we ask is $\sum_{i=1}^m \big( O(k_i \lg n) + O(\lg k_i) + O(k_i) \big) = O(n \lg n)$

Note, however, that each matrix question has $O(|D|) = O(n)$ tuples of $n$ variables each and therefore requires $O(n^2)$ time to generate. If we limit the number of tuples per question to a constant number, then we increase the number of questions asked to $\Omega(n^2)$.

\begin{lemma}
\label{lemma:constant}
 $\Omega(n^2)$ membership questions, with a constant number of tuples each, are required to learn existential expressions.
\end{lemma}

\emph{Proof:} Consider the class of queries on $n$ variables such that all variables in the set $X - \{x_i, x_j\}$ are body variables and the pair $x_{i}, x_{j}$ are head variables. Thus, the target class for our learning algorithm is the set of queries of this form:

\[\exists C_{ij} \rightarrow x_i \wedge \exists (C_{ij}\wedge x_i) \wedge \exists C_{ij} \rightarrow x_j \wedge \exists (C_{ij}\wedge x_j)\]

where $C_{ij} = X - \{x_i, x_j\}$ and $\exists (C_{ij}\wedge x_i), \exists (C_{ij}\wedge x_j)$ are guarantee clauses.

Any algorithm that learns such a query needs to determine exactly which of the possible $\binom{n}{2}$ pairs of variables in $n$ is the head variable pair. If we only have a constant number $c$ of tuples to construct a question with, we need to choose tuples that provide us with the most information. 

We can classify tuples as follows:

\emph{Class 1: Tuples where all variables are true}. Any question with such a tuple provides us with no information as the user will always respond that the set is an answer, regardless of the variable assignments in the other tuples.

\emph{Class 2: Tuples where one variable is false}. A question with one such tuple is always a non-answer as there are no tuples that satisfy $\exists (C_{ij} \wedge x_i) \wedge \exists (C_{ij}\wedge x_j)$. We denote a tuple where $x_i$ is false as $T_i$. A question with two or more class-2 tuples is an answer if it has two tuples $T_i, T_j$ such that $x_i, x_j$ are the head variables. If such a question is a non-answer, then for all the tuples $T_i, T_j$ in the question, the pair of variables $x_i, x_j$ are not the pair of head variables in the target query.

\emph{Class 3: Tuples where more than one variable is false}. If a question has \emph{only} class-3 tuples, then it will always be a non-answer. We denote a tuple where at least $x_i, x_j$ are false as $T_{ij}$. If $x_i, x_j$ are head variables then we need two more tuples $T_i$ and $T_j$ for the question to be an answer.  We cannot augment the question with more class-3 tuples to change it to an answer. Any other $T'_{ij}$ tuple, where $T'_{ij} \neq T_{ij}$ is bound to have at least one of its false $x'_i, x'_j$ variables as body variables: this makes $C_{ij}$ false, thus violating the clause $\exists (C_{ij} \wedge x_i) \wedge \exists (C_{ij}\wedge x_j)$.  
So questions with only class-3 tuples will always be non-answers and we improve the information gain of a question with some class-3 tuples by replacing those tuples with class-2 tuples.

Therefore, with $c$ tuples per question, we gain the most information from questions with only class-2 tuples. With $c \geq 2$, we construct a question with $c$ tuples such that for each variable $x_i \in H$, $H \subset X$ and $|H| = c$ the questions contains tuple $T_i$. If the question is an answer, then we narrowed our search for the pair of head variables to the set $H$. If the question is a non-answer, then we are sure that all pairs of $x_i, x_j$ variables with tuples $T_i, T_j$ in the question do not form the pair of head variables, so we eliminated $\binom{c}{2}$ pairs from consideration as head variables. An adversary will always respond to such questions with `non-answer'. In the worst case, we have to ask $\frac{\binom{n}{2}}{\binom{c}{2}} \approx \frac{n^2}{c^2} = \Omega(n^2)$ questions. \qed 

\subsection{Learning role-preserving qhorn}
\label{sec:rpqhornlearn}
Since some queries are more complex than others within the role-preserving qhorn query class it is natural to allow our learning algorithm more time, more questions and more tuples per question to learn the more complex target queries.  One can argue that such a powerful learning algorithm may not be practical or usable as it may ask many questions with many tuples each. If we assume that user queries tend to be simple (i.e they are small in size $k$ \cref{def:size} and have low causal densities $\theta$ \cref{def:cd}), then such an algorithm can be effective in the general case. 

Role-preserving qhorn queries contain two types of expressions: universal Horn expressions ($\forall x_1x_2 ... \rightarrow h$) and existential conjunctions ($\exists x_1x_2...$) (\cref{fig:rpqhornlabels} describes role-preserving qhorn terminology). In this section, we show that we can learn all universal Horn expressions with $O(n^{\theta + 1})$ questions and all existential conjunctions with $O(kn\lg n)$ questions. We show lower bounds of $\Omega(\frac{n}{\theta}^{\theta - 1})$ for learning universal Horn expressions and $\Omega(nk)$ for learning existential conjunctions. Since run-time is polynomial in the number of questions asked, our run-time is $poly(nk)$ and $poly(n^\theta)$ respectively. By setting a constant upper limit on the causal density of a head variable we can learn role-preserving qhorn queries in $poly(nk)$ time. 

\begin{figure}[htbp]
   \centering
   \includegraphics[width=0.65\linewidth]{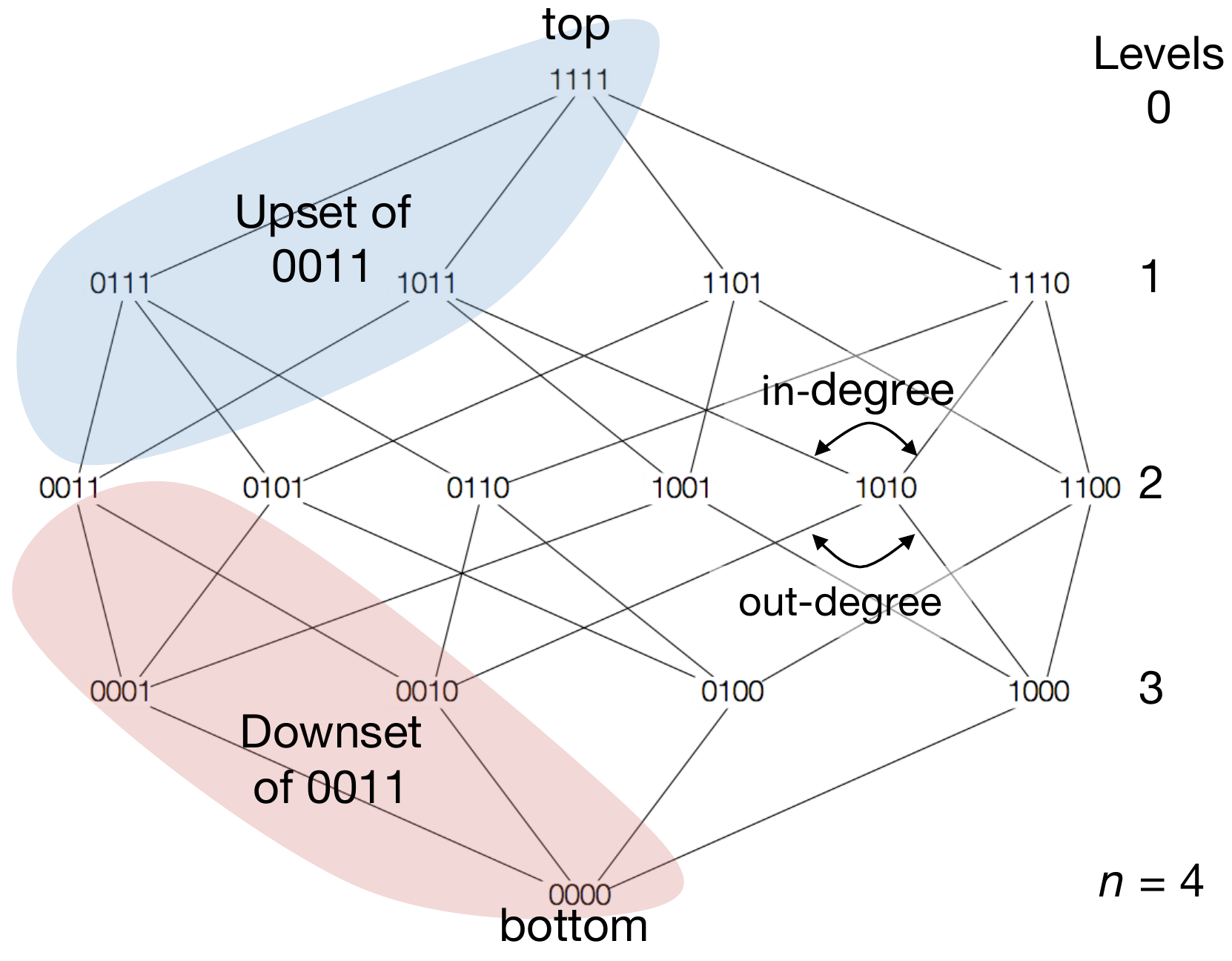}
   \caption[The Boolean lattice on four variables]{The Boolean lattice on four variables.}
   \label{fig:lattice}
\end{figure}

We employ a Boolean lattice on the $n$ variables of a query to learn the query's expressions. \cref{fig:lattice} illustrates the Boolean lattice and its key properties. Each point in the lattice is a tuple of true or false assignments to the variables. A lattice has $n+1$ levels. Each level $l$ starting from level 0 consists of tuples where exactly $l$ variables are false. A tuple's children are generated by setting exactly one of the true variables to false. Tuples at $l$ have out-degree of $n-l$, i.e. they have $n-l$ children and in-degree of $l$ or $l$ parents. A tuple has an \emph{upset} and a \emph{downset}. These are visually illustrated in \cref{fig:lattice}. If a tuple is not in the upset or downset of another tuple, then these two tuples are \emph{incomparable}. 

The gist of our lattice-based learning algorithms is as follows: 
\begin{enumerate}[leftmargin=0.5cm]
\item We map each tuple in the lattice to a distinct expression. This mapping respects a certain \emph{generality} ordering of expressions. For example, the lattice we use to learn existential conjunctions maps the top tuple in the lattice to the most specific conjunction $\exists x_1x_2...x_n$; tuples in the level above the bottom of the lattice map to the more general conjunctions $\exists x_1, \ \exists x_2, \ ...,\  \exists x_n$ \cref{sec:conjunct}. The exact details of this mapping for learning universal Horn expressions and learning existential conjunctions are described in the following section. 
\item We search the lattice in a \emph{top-to-bottom} fashion for the tuple that \emph{distinguishes} or maps to the target query expression. The learning algorithm generates membership questions from the tuples of the lattice and the user's responses to these questions either \emph{prune} the lattice or \emph{guide} the search.
\end{enumerate}

\subsubsection{Learning universal Horn expressions}
\label{sec:new-horn}

\begin{figure}[htbp]
   \centering
   \includegraphics[width=1\linewidth]{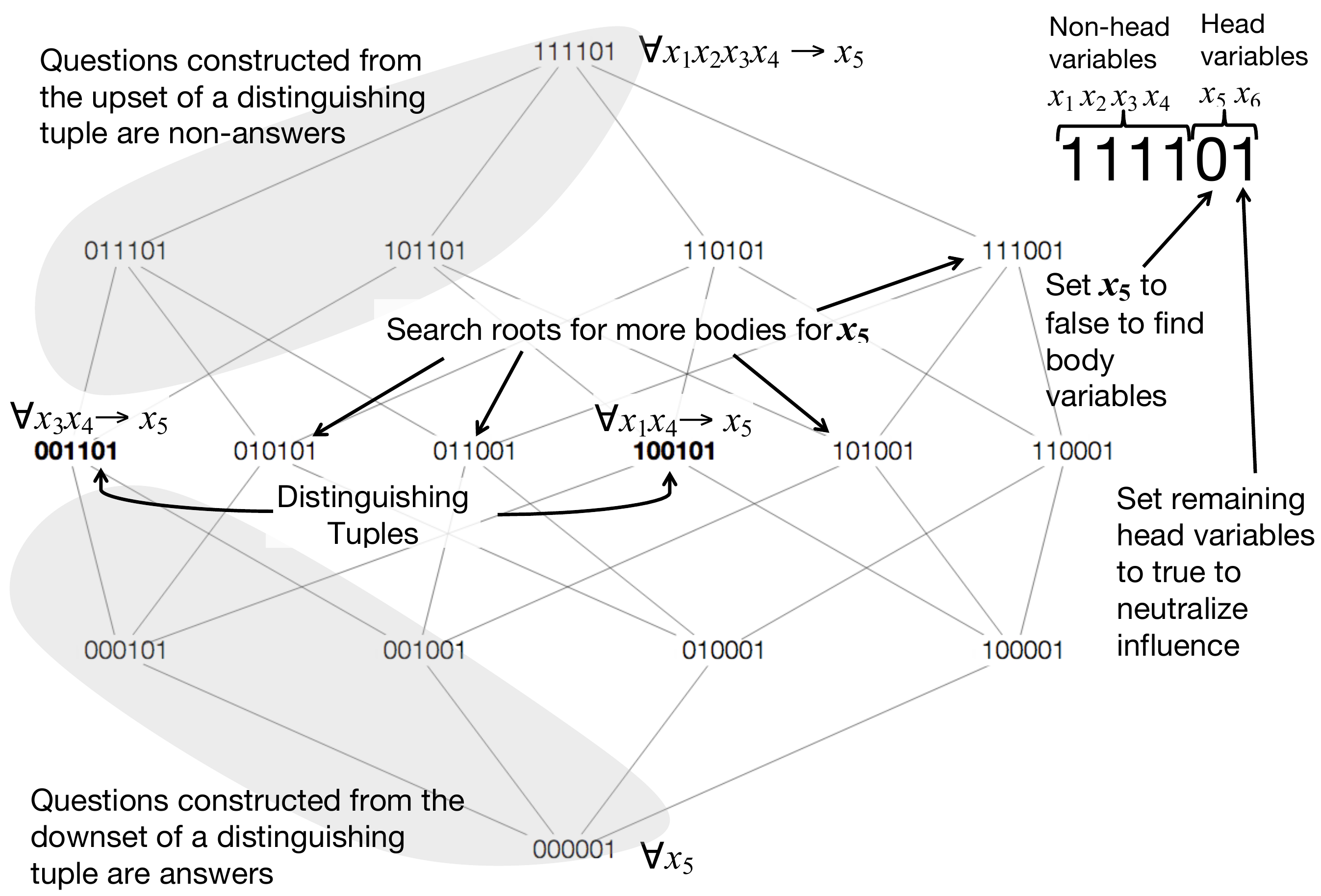}
   \caption{Learning bodies for a given head variable} 
   \label{fig:horn-learning}
\end{figure}

We first determine head variables of universal Horn expressions. We use the same algorithm of \cref{sec:qtypes}. The algorithm uses $O(n)$ questions. We then determine \emph{bodyless} head variables. To determine if $h$ is bodyless, we construct a question with two tuples: $1^n$ and a tuple where $h$ and all existential variables are false and all other variables are true. If the question is a non-answer then $h$ is bodyless. If $h$ is not bodyless then we utilize a special lattice (\cref{fig:horn-learning}) to learn $h$'s different bodies. In this lattice, we neutralize the effect of other head variables by fixing their value to true and we fix the value of $h$ to false.

\begin{defin}
\label{def:uni-distinguish}
A universal Horn expression for a given head variable $h$ is \textbf{distinguished} by a tuple if the true variables of the tuple represent a complete body for $h$. 
\end{defin}
Thus, each tuple in the lattice distinguishes a unique universal Horn expression. For example, consider the target query:
\begin{equation*}
\begin{array}{l}
\forall x_1x_4 \rightarrow x_5 \ \forall x_3x_4 \rightarrow x_5\  \forall x_1x_2 \rightarrow x_6 \\
\exists x_1x_2x_3 \ \exists x_2x_3x_4 \ \exists x_1x_2x_5 \ \exists x_2x_3x_5x_6 \\
\end{array}
\end{equation*}

In the target query, the head variable $x_5$ has two universal Horn expressions:
\[\forall x_1x_4 \rightarrow x_5 \ \forall x_3x_4 \rightarrow x_5\]
In \cref{fig:horn-learning}, we marked the two tuples that \emph{distinguish} $x_5$'s universal Horn expressions: 100101 and 001101. Notice that the universal Horn expressions are ordered from most to least specific. For example the top tuple of the lattice in \cref{fig:horn-learning} is the distinguishing tuple for the expression $\forall x_1x_2x_3x_4 \rightarrow x_5$. While the bottom tuple is the distinguishing tuple for the expression $\forall x_5$. Our learning algorithm searches for distinguishing tuples of only \emph{dominant} universal Horn expressions.  

A membership question with a distinguishing tuple and the all-true tuple (a tuple where all variables are true) is a non-answer for one reason only: \emph{it violates the universal Horn expression it distinguishes}.  This is because the all-true tuple satisfies all the other expressions in the target query and the distinguishing tuple sets a complete set of body variables to true but the head to false. More importantly, all such membership questions constructed from tuples in the upset of the distinguishing tuple are non-answers and all questions constructed from tuples in the downset of the distinguishing tuple are answers. Thus, the key idea behind the learning algorithm is to efficiently search the lattice to find a tuple where questions constructed from tuples in the upset are non-answers and questions constructed from tuples in the downset are answers.

Given a head variable $h$ and $n$ (non-head) variables, we use a Boolean lattice on $n$ variables (with $h=0$ and all other head variables set to true). We construct a membership question with a tuple $t$ from the lattice and the all-true tuple --- a tuple where all variables, including the head variable, are true. We begin by describing how we can use the lattice to find just one set of body variables that determine $h$ with $O(n)$ questions. We start at the top of the lattice, we construct a question from the top tuple and proceed as follows:
\begin{enumerate}[leftmargin=0.5cm]
\item If the question is an answer, then it does not contain an entire set of body variables that determine $h$. We prune its \emph{downset}. We move to the next tuple on the same level of the lattice.
\item If the question is a non-answer then some of the true variables in $t$ form a body and we move down the lattice (skipping previously pruned tuples). If all of $t$'s children are answers, then $t$ is a \emph{universal distinguishing tuple} for the head variable $h$.
\end{enumerate}

This lattice-based algorithm is equivalent to the simple procedure listed in Algorithm \ref{alguni}.
\begin{algorithm}[htb]                      
\small
\caption{\small Learn one body for head variable $h$}          
\label{alguni}                           
\begin{algorithmic}                    
    \State $h$: the head variable we are learning a body for
    \State $N$: the set of all non-head variables
    \State $H$: the set of all head variables
    \State
    \State $X \leftarrow \emptyset$ \Comment{\small\emph{$X$ holds all variables that are not in the body.}}
    \For{$x \in N$}
    	\State SetTuple$(t_0, N \cup H, \mathbf{T})$
    	\Comment{\small\emph{This is the all-true tuple.}}
		\State $t_1 \leftarrow t_0$
		\State SetTuple$(t_1, X \cup \{x, h\}, \mathbf{F})$
		\State isAnswer $\leftarrow$ Ask($\{t_0, t_1\}$)
		\If{\textbf{not} isAnswer}
			\State $X\leftarrow X \cup x$
		\EndIf
	\EndFor
	\State \Return $N - X$
\end{algorithmic}
\end{algorithm}

Once we find a body, we can safely eliminate its \emph{upset}. Any body in the upset is \emph{dominated} \ref{ax:uni-dominate} by the discovered distinguishing tuple. Looking at \cref{fig:horn-learning}, we notice that the upset simply contains all tuples where all body variables of the distinguishing tuple are true. The remaining lattice structure is rooted at tuples where one of the body variables is false. Since two incomparable bodies need to differ on at least one body variable, we set one body variable to false and search the resulting sub-lattices for bodies.

\begin{theorem}
$O(n^\theta)$ membership questions, where $\theta$ is the causal density of the given head variable $h$, are sufficient to the learn the $\theta$ universal Horn expressions of $h$.
\end{theorem}
\emph{Proof:} Let $b_i$ denote the number of body variables for each distinguishing tuple $t_i$ found. Initially we set $b_0$ to $n$ and we search the entire lattice or the $n$ sub-lattices rooted at the tuples where exactly one Boolean variable is false. In \cref{fig:horn-learning} those are the tuples at level 1: $\{0111\emph{01},$ $1011\emph{01},$ $1101\emph{01},$ $1110\emph{01}\}$. 

If the first distinguishing tuple found has $|B_1|$ true variables, then we need to search $|B_1|$ sub-lattices for bodies.  For example, after finding the distinguishing tuple $0011\emph{01}$, we continue searching for more distinguishing tuples from $|B_1| = 2$ roots: $\{1101\emph{01}, 1110\emph{01}\}$. 

Suppose we find a second distinguishing tuple: $1001\emph{01}$ with $B_2$ body variables; then we need to search for more bodies in the sub-lattices rooted at tuples where one of each body variable from the distinct bodies are set to false. Our new $|B_1| \times |B_2|$ roots are: $\{0101\emph{01}, \ 0110\emph{01}, \ 1010\emph{01}, \  1110\emph{01}\}$. These \emph{search roots} are illustrated in \cref{fig:horn-learning}.

In the worst case, we ask $O(n)$ questions to find a body. Thus to determine all $\theta$ expressions for a universal head variable, an upper bound on the number of questions, $Q$, is:
\[ Q \leq \begin{array}{l} (n) + (|B_1| + n) + (|B_1| \times |B_2| + n) + ... + \\ 
(|B_1| \times |B_2| \times ... \times |B_\theta|)\end{array}\]
\[Q \leq n\theta + \sum_{b=1}^{\theta}(\prod_{i=1}^{b}|B_i|) \leq n\theta + \sum_{i=1}^{\theta}(n^i) = O(n^\theta) \qed \]

Since there are $O(n)$ head variables and for each head variable we ask $O(n^\theta)$ questions to determine its universal Horn expressions, we learn all universal Horn expression with $O(n \times n^\theta) = O(n^{\theta + 1})$ questions.

\begin{theorem}
$\Omega((\frac{n}{\theta})^{\theta-1})$ membership questions, where $\theta$ is causal density of $h$, are required to the learn the $\theta$ universal Horn expressions of $h$.
\end{theorem}

Consider the class of role-preserving queries with $\theta$ universal Horn expressions and $n$ body variables where each expression $C_i$ for $1 \leq i < \theta$ consists of $\frac{n}{\theta-1}$ body variables, $B_i$, that determine $h$.  The expression $C_{\theta}$ consists of $n - \theta - 1$ body variables, $B_{\theta}$, such that $|B_{\theta} \cap B_i| = \frac{n}{\theta-1} - 1$.

The following is an example instance with $n = 12$ body variables and $\theta = 4$ expressions: 
\begin{equation*}
\begin{array}{l}
\forall x_1x_3x_5x_9 \rightarrow h \ \forall x_2x_4x_6x_{10} \rightarrow h \ \forall x_7x_8x_{11}x_{12} \rightarrow h \\
\forall x_1x_2x_3x_4x_7x_8x_9x_{10}x_{11} \rightarrow h
\end{array}
\end{equation*}

If we construct a tuple where two or more variables from each body $B_i$ and the head variable are false, then all $\theta-1$ bodies and $B_{\theta}$ are not satisfied and a question consisting of such a tuple and the all-true tuple ($1^{n+1}$) will always be an answer. 

Alternatively, if we set all of variables of one body $B_i$ to true and the head variable to false then a question with such a tuple will always be a non-answer. Therefore, we can only set exactly one variable from each body to false to learn $B_{\theta}$. There are $\frac{n}{\theta-1}$ choices per body for which body variable to be false, leaving us with $(\frac{n}{\theta-1})^{\theta-1}$ possible questions. If a question is an answer, we eliminate only one combination of body variables for $B_{\theta}$ and if the question is a non-answer then $B_{\theta}$ consists of all body variables that are true. In the worst-case the user responds that each question is an answer forcing the algorithm to ask $(\frac{n}{\theta-1})^{\theta-1} - 1 = \Omega((\frac{n}{\theta})^{\theta-1})$ questions. \qed

\subsubsection{Learning existential conjunctions}
\label{sec:conjunct}

To learn existential conjunctions of a query we use the full Boolean lattice on all $n$ variables of a query (including head variables).

\begin{defin}
\label{def:exi-distinguish}
An existential conjunction $C$ is \textbf{distinguished} by a tuple if the true variables of the tuple are the variables of the conjunction.
\end{defin}

Thus, each tuple in the lattice distinguishes a unique existential conjunction. For example, consider the target query:
\begin{equation*}
\begin{array}{l}
\forall x_1x_4 \rightarrow x_5 \ \forall x_3x_4 \rightarrow x_5\  \forall x_1x_2 \rightarrow x_6 \\
\exists x_1x_2x_3 \ \exists x_2x_3x_4 \ \exists x_1x_2x_5 \ \exists x_2x_3x_5x_6 \\
\end{array}
\end{equation*}
The conjunction $\exists x_2x_3x_5x_6$ is distinguished by the tuple $011011$ in a six-variable Boolean lattice.

Existential conjunctions are ordered from most to least specific on the lattice. For example, the top tuple $111111$ of a six-variable lattice is the distinguishing tuple for the expression $\exists x_1x_2x_3x_4x_5x_6$; the tuples $\{00001, 000010, 000100, 001000, 010000, 100000\}$ at level five of the lattice are the distinguishing tuples for the expressions $\exists x_6, \ \exists x_5, \ \exists x_4, \ \exists x_3, \ \exists x_2, \ \exists x_1$ respectively.

Our learning algorithm searches for distinguishing tuples of a \emph{normalized} target query. For example, the target query above is normalized to the following semantically equivalent query using \ref{ax:exi-extend}:
\begin{equation}
\label{query:tq}
\begin{array}{l}
\forall x_1x_4 \rightarrow x_5 \ \forall x_3x_4 \rightarrow x_5\  \forall x_1x_2 \rightarrow x_6 \\
\exists x_1x_2x_3x_6 \ \exists x_2x_3x_4x_5 \ \exists x_1x_2x_5x_6 \ \exists x_2x_3x_5x_6 \\
\end{array}
\end{equation}
This query has the following \emph{dominant} conjunctions (which include guarantee clauses): 
\[\exists x_1x_4x_5 \ \exists x_1x_2x_3x_6 \ \exists x_2x_3x_4x_5 \ \exists x_1x_2x_5x_6 \ \exists x_2x_3x_5x_6\]

A membership question with \textbf{all dominant} distinguishing tuples of a query is an answer: all existential conjunctions (including guarantee clauses) are satisfied. For example, a question with the tuples: $\{100110, 111001, 011110, 110011, 011011\}$ is an answer for the target query above (\ref{query:tq}).

Replacing a distinguishing tuple with its children results in a non-answer: the existential conjunction of that tuple is no longer satisfied. For example replacing $011011$ with its children $\{001011, 010011, 011001, 011010\}$ results in a membership question where none of the tuples satisfy the expression $\exists x_2x_3x_5x_6$.

Replacing a distinguishing tuple with any tuple in its upset \emph{that does not violate a universal Horn expression} still results in an answer.

Thus, the learning algorithm searches level-by-level from top-to-bottom for distinguishing tuples by detecting a change in the user's response to a membership question from answer to non-answer. The efficiency of the learning algorithm stems from \emph{pruning}: when we replace a tuple with its children, we prune those down to a minimal set of tuples that still dominate all the distinguishing tuples.

We describe the learning algorithm (Alg.~\ref{exilearnalg}) with an example and then prove that the learning algorithm runs in $O(k n \lg n)$ time (\cref{thm:cost}). We also prove the algorithm's correctness (\cref{thm:correct}) and provide a lower bound of $O(nk)$ for learning existential conjunctions (\cref{thm:lb}).

\begin{algorithm}[htb]                      
\caption{\small Find Existential Distinguishing Tuples}          
\label{exilearnalg}                           
\small
\begin{algorithmic}                    
    \State $T \leftarrow \{1^n\}$ \Comment{\small\emph{The top tuple.}}
    \State $D \leftarrow \{\}$ \Comment{\small\emph{$D$ is the set of discovered distinguishing tuples.}}
    \While{$T \neq \emptyset$}
    \State $T' \leftarrow \{ \}$
    \For{$t \in T$}
		\State $C \leftarrow \mathrm{Children}(t)$
		\State $C \leftarrow \mathrm{Remove Universal Horn Violations}(C)$
		\State $T \leftarrow T - \{t\}$
		\State isAnswer $\leftarrow$ Ask($D \cup T \cup C \cup T'$)
		\If{isAnswer}
			\State $T' \leftarrow T' \cup \mathrm{\textbf{Prune}}(C, T \cup D)$
		\Else
			\State $D \leftarrow D \cup \{t\}$
		\EndIf
    \EndFor
    \State $T \leftarrow T'$
    \EndWhile
    \State \Return $D$
\end{algorithmic}
\end{algorithm}

\begin{algorithm}[htb]                      
\caption{\small Prune}          
\label{alg2}
\small                           
\begin{algorithmic}                    
    \State $T$: {\small the tuples to prune}
    \State $O$: {\small other tuples}
    \State $K \leftarrow \{ \}$  \Comment{\small\emph{$K$ is the set of tuples to keep.}}	
    \State Split $T$ into $T_1$ (1$^{st}$ half) and  $T_2$ (2$^{nd}$ half).
    \While{$T_1 \cup T_2 \neq \emptyset$}
    \State isAnswer $\leftarrow$ Ask($T_1 \cup K \cup O$)	
    \If{isAnswer}
    	\State Split $T_1$ into $T_1$ (1$^{st}$ half) and $T_2$ (2$^{nd}$ half).
    \Else
    	\If{$|T_2| = 1$}
			\State $K \leftarrow K \cup T_2$
		\Else
			\State Add 1$^{st}$ half of $T_2$ to $T_1$. Set $T_2$ to 2$^{nd}$ half of $T_2$.
    	\EndIf
    \EndIf
    \EndWhile
    \State \Return $K$
\end{algorithmic}
\end{algorithm}

Suppose we wish to learn the existential conjunctions of the target query listed in (\ref{query:tq}). We use the six-variable Boolean lattice with the following modification: we remove all tuples that violate a universal Horn expression. These are tuples where the body variables of a universal Horn expression are true and the head variable is false. For example, the tuple $111110$ violates $\forall x_1x_2 \rightarrow x_6$ is therefore removed from the lattice.

\noindent\textbf{Level 1:} We start at the top of the lattice. Since the tuple $111111$ will satisfy any query, we skip to level one. We now construct a membership question with all the tuples of level 1 (after removing the tuples that violate universal Horn expressions: $111110, 111101$): $111011, 110111, 101111, 011111$. If such a question is a non-answer, then the distinguishing tuple is one level above and the target query has one dominant existential conjunction: $\exists x_1x_2x_3x_4x_5x_6$. 

\noindent\includegraphics[width=1\linewidth]{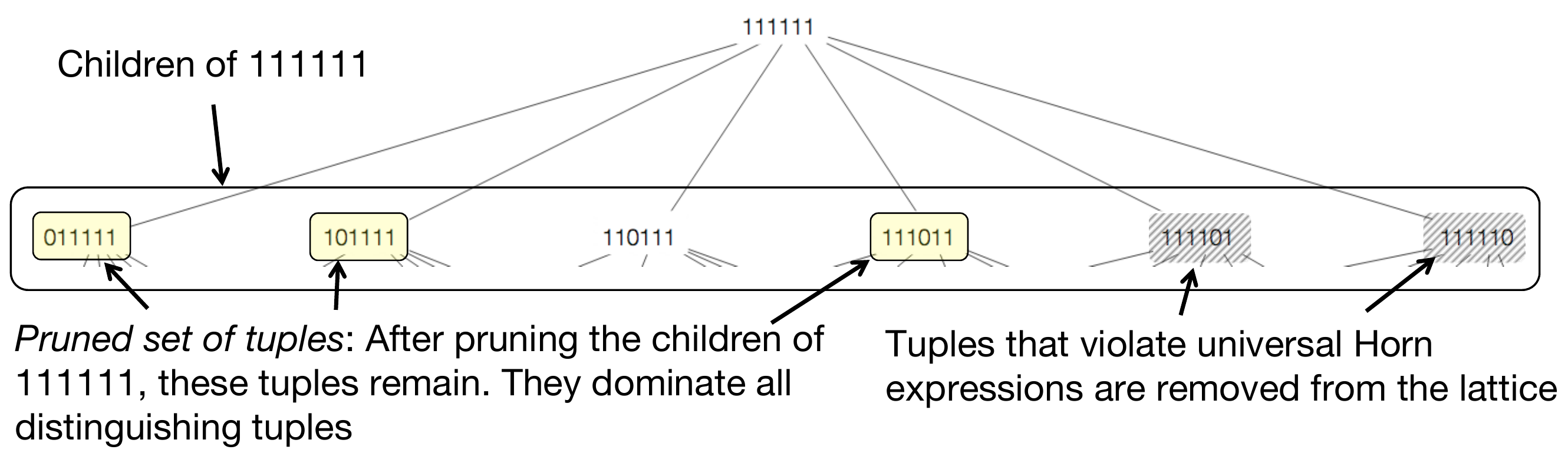}

If the question is an answer, we need to search for tuples we can safely \emph{prune}. So we remove one tuple from the question set and test its membership. Suppose we prune the tuple $110111$, the question is still an answer since all conjunctions of the target query are still satisfied: the remaining set of tuples still dominate the distinguishing tuples of the target query.

We then prune $011111$. This question is a non-answer since no tuple satisfies the clause $\exists x_2x_3x_4x_5$. We put $011111$ back in and continue searching at level one for tuples to prune. We are left with the tuples: $111011$, $101111$ and $011111$. Note that we asked $O(n)$ questions to determine which tuples to safely prune. We can do better. In particular, we only need $O(\lg n)$ questions for each tuple we need to keep if we use a binary search strategy. 

\noindent\textbf{Level 2:}  We replace one of the tuples, $111011$, with its children on level 2: $\{011011, 101011, 110011, 111001\}$. Note, that we removed $111010$ because it violates $\forall x_1x_2 \rightarrow x_6$. As before we determine which tuples we can safely prune. We are left with $\{110011, 111001\}$. 

\noindent\includegraphics[width=1\linewidth]{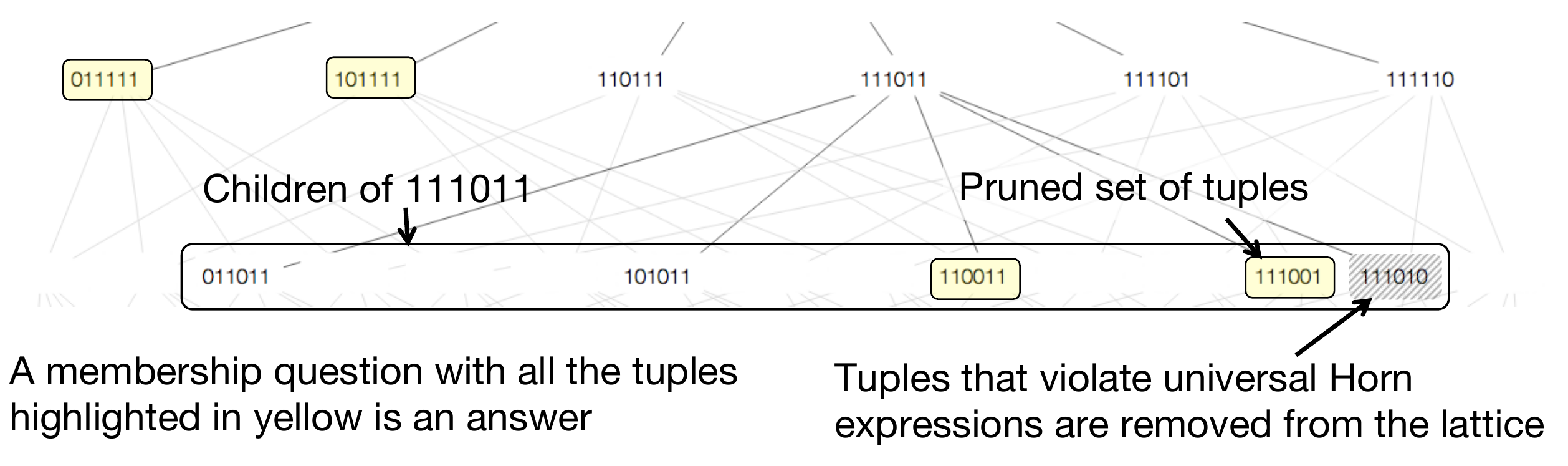}

Similarly we replace $101111$ with its children on level 2: $\{001111, 100111, 101011, 101110\}$. We did not consider $101101$ because it violates $\forall x_3x_4 \rightarrow x_5$. We can safely prune the children down to one tuple: $101110$. We then replace $011 111$ with its children on level 2 and prune those down to $\{011011, 011110\}$. At the end of processing level $2$, we are left with the tuples: $\{110011, 111001, 101110, 011011, 011110\}$. 
We repeat this process again now replacing each tuple, with tuples from level $3$.

\noindent\textbf{Level 3:}  When we replace $011110$ with its children $\{010110, 011010, 001110\}$, we can no longer satisfy $\exists x_2x_3x_4x_5$. The question is a non-answer and we learn that $011110$ is a distinguishing tuple and that $\exists x_2x_3x_4x_5$ is a conjunction in the target query. Note that we did not consider the child tuple $011100$ because it violates the universal Horn expression $\forall x_3x_4 \rightarrow x_5$. We fix $011110$ in all subsequent membership questions.

\noindent\includegraphics[width=1\linewidth]{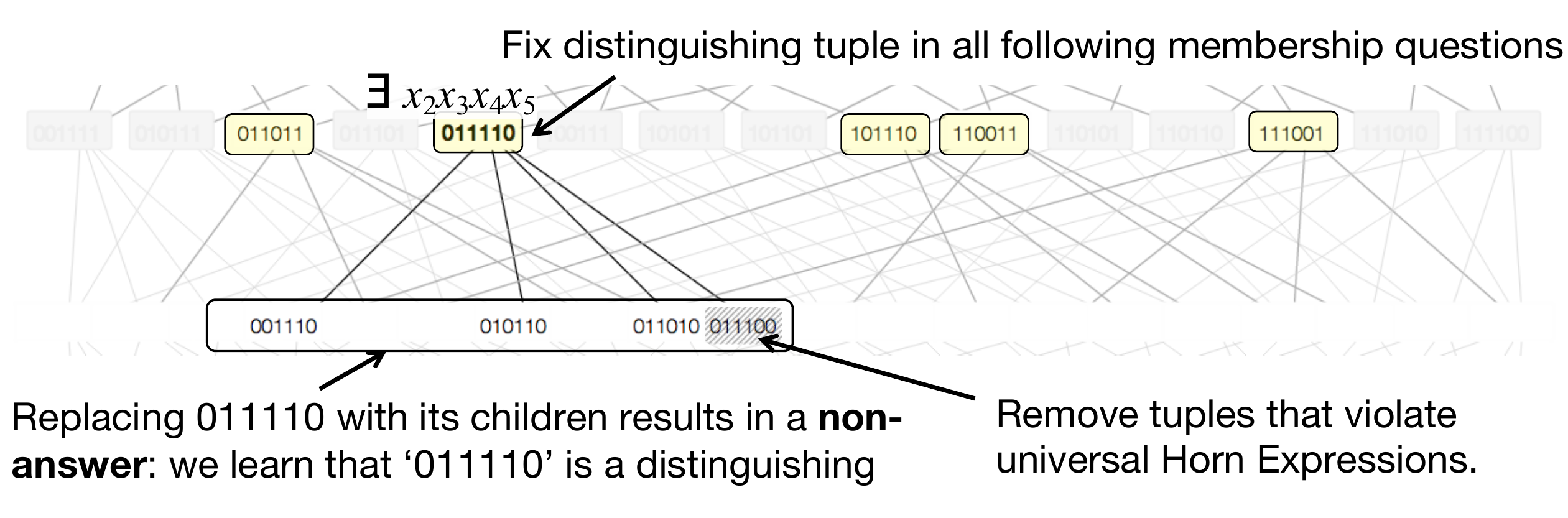}

When we replace $011011$ with its children $\{001011, 010011, 011001, 011010\}$, we can no longer satisfy $\exists x_2x_3x_5x_6$. The question is a non-answer and we learn that $011011$ is a distinguishing tuple and that $\exists x_2x_3x_5x_6$ is a conjunction in the target query. We fix $011011$ in all subsequent membership questions.

When we replace $111001$ with its children $\{011001, 101001, 110001\}$, the question is a non-answer, and we learn that $111001$ is distinguishing tuple and that $\exists x_1x_2x_3x_6$ is a conjunction in the target query. Note that we did not consider the tuple $111000$ because it violates $\forall x_1x_2 \rightarrow x_6$. We fix $111001$ in all subsequent membership questions.

We can replace $101110$ with the children $\{001110, 100110, 101010\}$. Note that the child $101100$ is removed because it violates $\forall x_1x_4 \rightarrow x_5$. We can safely prune the children down to one tuple $100110$.

When we replace $110011$ with its children $\{010011, 100011, 110001\}$, we can no longer satisfy $\exists x_1x_2x_5x_6$.  Thus, the question is a non-answer and we learn that $110011$ is a distinguishing tuple. Note that we did not consider the tuple $110010$ because it violates $\forall x_1x_2 \rightarrow x_6$. 

At this stage, we are left with the following tuples: \[\{\mathbf{110011}, 100110, \mathbf{111001}, \mathbf{011011}, \mathbf{011110}\}\]

At this point, we can continue searching for conjunctions in the downset of 100110 which is the distinguishing tuple for a known guarantee clause for the universal Horn expression: $\forall x_1x_4\rightarrow x_5$. As an optimization to the algorithm, we do not search the downset because all tuples in the downset are dominated by 100110\footnote{We can relax the requirement of guarantee clauses for universal Horn expressions and our learning algorithms will still function correctly if they are allowed to ask about the membership of an empty set.}.

\noindent\includegraphics[width=1\linewidth]{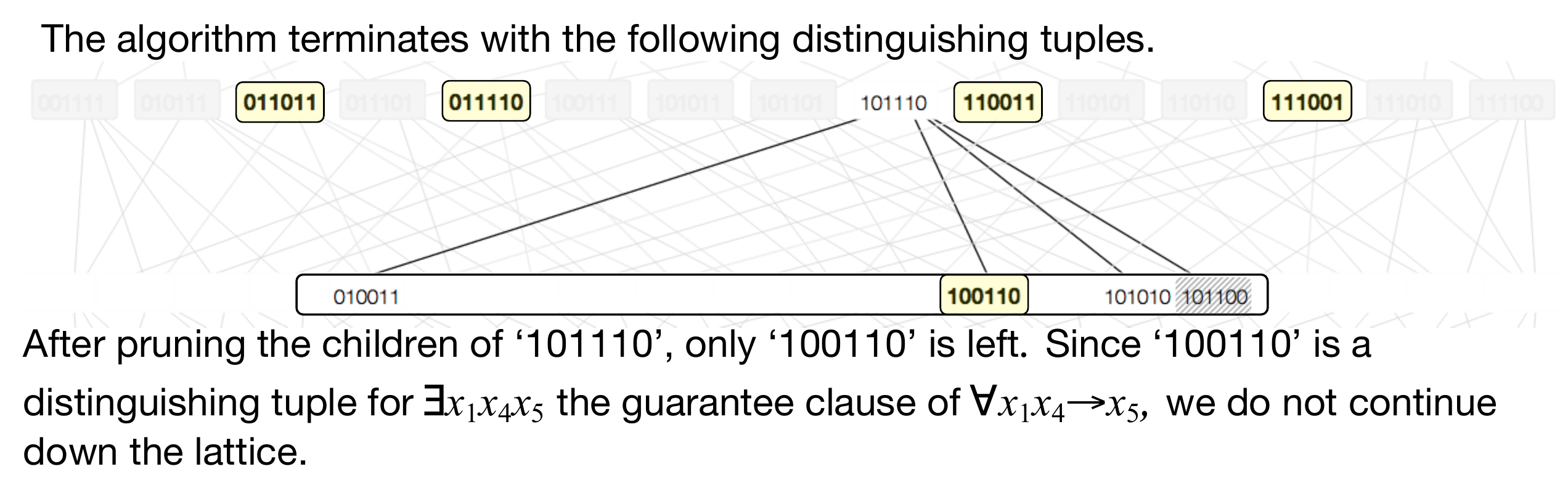}

The learning algorithm terminates with the following distinguishing tuples $\{110011, 100110, 111001, 011011, 011110\}$ which represent the expressions: 
\[\exists x_1x_2x_5x_6 \ \exists x_1x_4x_5 \ \exists x_1x_2x_3x_6 \ \exists x_2x_3x_5x_6 \ \exists x_2x_3x_4x_5\] 

\begin{theorem}
\label{thm:correctness}
The lattice-based learning algorithm finds the distinguishing tuples of all dominant existential conjunctions of a normalized target query.
\end{theorem}

The algorithm will find a dominant distinguishing tuple if there exists a path from at least one tuple in its current pruned set of tuples to the distinguishing tuple. A path between two tuples $t_0, t_1$ is simply the sequence of variables to set to false to get from $t_0$ to $t_1$. When pruning, we ensure that the pruned set of tuples \emph{dominates} all the distinguishing tuples: so there always exists a path from at least one tuple in the pruned set of tuples to a distinguishing tuple provided our lattice is complete. 

To see that removing tuples that violate universal Horn expressions will not impact the existence of a path, suppose a tuple $t_a$ in the pruned set dominates a distinguishing tuple $t_d$ and $t_a$ does not violate any universal Horn expressions. We relabel all the non-head variables in $t_a$ to $e^a_0 ... e^a_m$ and all the head variables to $h^a_0 ... h^a_{n-m}$. Similarly, we relabel the non-head variables in $t_d$ with $e^d_0 ... e^d_m$ and the head variables with $h^d_0 ... h^d_{n_m}$. Consider the tuple $t_b$ which has the values of $e^d_0 ... e^d_m$ for its non-head variables and the values of $h^a_0 ... h^a_{n-m}$ for its head variables. Clearly there exists a path from tuple $t_a$ to tuple $t_b$ that does not encounter any tuples that violate universal Horn expressions because the head variables in both $t_a$ and $t_b$ have the same values. 

Similarly, there exists a path from $t_b$ to $t_d$. Since $t_b$ is in the upset of $t_d$, $h^i_0 ... h^i_{n-m}$ dominates $h^d_0 ... h^d_{n-m}$. In a normalized query, all existential conjunctions are expanded to include head variables that are implied by the variables of a conjunction \ref{ax:exi-extend}. Thus, $t_d$ does not violate any universal Horn expression and the path from $t_b$ to $t_d$ only sets head variables to false that do not violate any universal Horn expressions.

Suppose the learning algorithm favors another tuple $t_c$ (instead of $t_b$) such that $e^c_0 ... e^c_{m}$ is in the downset of $e^a_0 ... e^a_{m}$ and in the upset of $e^d_0 ... e^d_{m}$ and $h^c_0 ... h^c_{n-m}$ is in the downset of $h^a_0 ...h^a_{n-m}$ and in the upset of $h^d_0 ... h^d_{n-m}$. If the learning algorithm reaches $t_c$ then $t_c$ does not violate any universal Horn expressions and by induction (let $t_a = t_c$) there also exists a path from $t_c$ to $t_d$. \qed

\begin{theorem}
\label{thm:cost}
The lattice-based learning algorithm asks $O(kn \lg n)$ membership questions where $k$ is the number of existential conjunctions. 
\end{theorem}

\emph{Proof:} 
Consider the cost of learning one distinguishing tuple $t_l$ at level $l$. From the top of the Boolean lattice to $t_l$, there is at least one tuple $t_i$ at each level $i$ ($0 < i < l$) that we did not prune and we traversed down from to get to $t_l$. Let $N_i$ be the set of $t_i$'s siblings. At each level $i$, we asked at most $\lg |N_i|$ questions. $|N_i| = n - (i-1)$ or the out-degree of $N_i$'s parent. In the worst-case, $l = n$, and the cost of learning $t_l$ is $\sum_{i = 1}^{n}\lg (n - (i-1)) \leq \sum_{i = 1}^{n}\lg n  = O(n\lg n)$. With $k$ distinguishing tuples we ask at most $O(kn\lg n)$ questions. \qed

\begin{theorem}
\label{thm:lb}
$\Omega(nk)$ questions are required to learn existential conjunctions.
\end{theorem}

From an information theoretic perspective, $\Omega(nk)$ questions is a lower bound on the number of questions needed to learn existential expressions. Consider level $n/2$ of the Boolean lattice, which holds the maximum number of non-dominated distinguishing tuples. There are $\binom{n}{n/2}$ tuples at this level. Suppose we wish to learn $k$ existential expressions at this level. There are $\binom{\binom{n}{n/2}}{k}$ possible $k$-expressions. Since each question provides a bit of information, a lower bound on the number of questions needed is $\lg\binom{\binom{n}{n/2}}{k}$.

\[\lg\binom{\binom{n}{n/2}}{k} \geq \lg\binom{2^{n/2}}{k} \geq \lg\Big(\frac{2^{n/2}}{k}\Big)^k = \frac{nk}{2} - k \lg k\]

\section{Query Verification}
\label{sec:verify}

\begin{figure*}[htbp]
   \centering
   \includegraphics[width=1\linewidth]{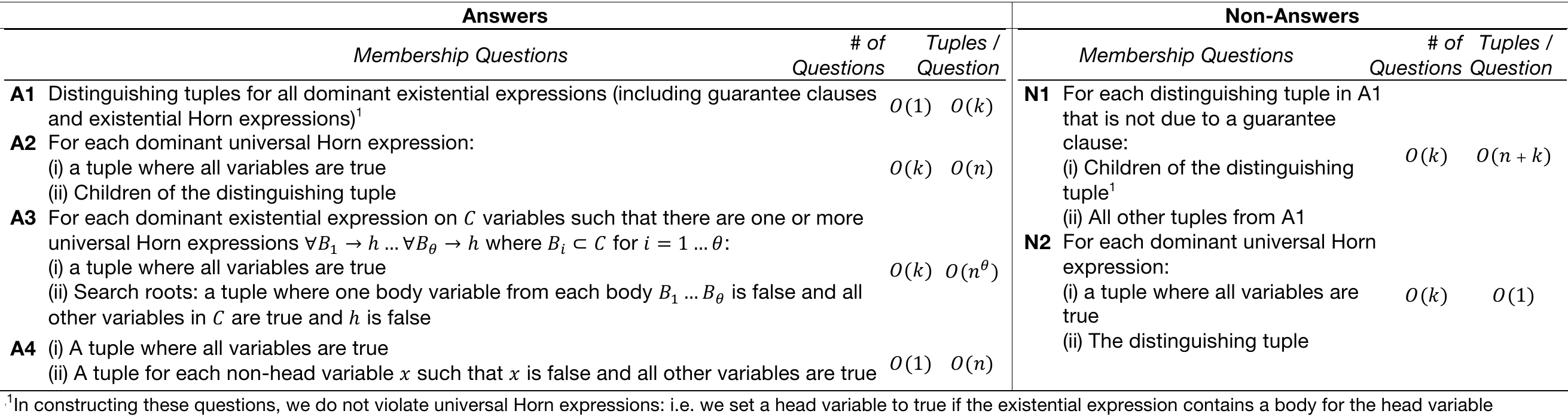}
   \caption{Membership questions of a verification set.}
   \label{fig:two-table}
\end{figure*}

A query verifier constructs a set of membership questions to determine whether a given query is correct. The verifier will not find an alternate query if the query is incorrect. Thus, while query learning is a \emph{search problem} --- a learner searches for the one correct query that satisfies the user's responses to membership questions; query verification is the \emph{decision problem} --- a verifier decides if a given query is correct or incorrect given the user's responses to membership questions. 

Our approach to query verification is straightforward: for a given role-preserving qhorn\footnote{Since qhorn-1 is a sub-class of role-preserving qhorn, our verification approach works for both query classes.} query $q_g$, we generate a \emph{verification set} of $O(k)$ membership questions, where $k$ is the number of expressions in $q_g$. Note that our learning algorithm for role-preserving qhorn queries asks $O(n^{\theta+1} + kn\lg n)$ questions. If the user's intended query $q_i$ is semantically different from the given query $q_g$, then for at least one of the membership questions $M$ in the verification set $q_g(M) \neq q_i(M)$. 

\begin{proposition}
\label{prop:diff}
A user's intended query $q_i$ is semantically different from a given query $q_g$ iff $q_i$ and $q_g$ have distinct sets of existential \cref{def:exi-distinguish} and universal \cref{def:uni-distinguish} distinguishing tuples.
\end{proposition}

Suppose we try to learn the two role-preserving qhorn queries $q_i$ and $q_g$. If $q_i$ and $q_g$ are semantically different, then our learning algorithm will terminate with distinct sets of existential\cref{def:exi-distinguish} and universal\cref{def:uni-distinguish} distinguishing tuples for each query. The verification set consists of membership questions that detect semantic differences between two queries by detecting differences in their respective sets of distinguishing tuples. \cref{fig:two-table} lists six types of membership questions from which the verification algorithm constructs a verification set for a given query.

We explain how to construct each question for an example query in \cref{sec:examplevs}.

\subsection{Normalizing User-specified Queries}

All membership questions are constructed from \emph{dominant} existential \ref{ax:exi-dominate} or universal \ref{ax:uni-dominate} expressions. A user-specified query $q_g$ may contain redundant or dominated expressions. For example $\exists x_1x_2$ is dominated by $\exists x_1x_2x_3$ \ref{ax:exi-dominate} and is therefore redundant. 

\subsubsection{Dominant Existential Distinguishing Tuples}
To find the dominant existential expressions a simple routine orders all existential expressions by the number of participating variables from largest to smallest. For each expression in the ordered list, we remove all other expressions in the list whose participating variables are a subset of the variables of the current expression. This leaves us with the set of dominant existential expressions. 

To construct a distinguishing tuple from an existential expression, we set all participating variables of the expression to true and the remaining variables to false. If setting one of the remaining variables to false violates a universal Horn expression, we set it to true. This is equivalent to rewriting the given query $\exists x_1x_2 \ \forall x_1 \rightarrow h$ to a semantically equivalent query $\exists x_1x_2h \ \forall x_1 \rightarrow h$ \ref{ax:exi-extend}.

\subsubsection{Dominant Universal Distinguishing Tuples}
A user-specified query $q_g$ may also contain redundant universal Horn expressions. For example $\forall x_1x_2 \rightarrow x_3$ is dominated by $\forall x_1 \rightarrow x_3$ \ref{ax:uni-dominate}. To find the dominant universal distinguishing expressions, a simple routine orders all universal Horn expressions by the number of the participating variables from smallest to largest. For each expression in the ordered list, we remove all other expressions in the list whose participating variables are a superset of the variables of the current expression. This leaves us the set of dominant universal Horn expressions. 

To construct a universal distinguishing tuple from a universal Horn expression, we set the head variable to false and all body variables of the expression to true. The remaining head variables are set to true and the remaining body variables are set to false.

\subsection{Example Verification Set}
\label{sec:examplevs}

We demonstrate the construction of a verification set on a role-preserving qhorn query with six Boolean variables $x_1, ..., x_6$. This is the same query that we previously learned in \cref{sec:conjunct}.
\begin{equation*}
\begin{array}{l}
\forall x_1x_4 \rightarrow x_5 \ \forall x_1x_2 \rightarrow x_6 \  \forall x_3x_4 \rightarrow x_5 \\
\exists x_1x_2x_3 \ \exists x_2x_3x_4 \ \exists x_1x_2x_5 \ \exists x_2x_3x_5x_6
\end{array}
\end{equation*}

\noindent\textbf{[A1]~~}For each existential expression and guarantee clause, we construct the following distinguishing tuples.

\begin{tabular}{lcll}
$\exists x_1x_2x_3$ & $\Rightarrow$ & 111001 & {\em Do not violate } $\forall x_1x_2 \rightarrow x_6$ \\
$\exists x_2x_3x_4$ & $\Rightarrow$ & 011110 & {\em Do not violate } $\forall x_3x_4 \rightarrow x_5$ \\
$\exists x_1x_2x_5$ & $\Rightarrow$ & 110011 & {\em Do not violate } $\forall x_1x_2 \rightarrow x_6$ \\
$\exists x_2x_3x_5x_6$&$\Rightarrow$& 011011 & \\
& & & {\em The guarantee clause of:} \\
$\exists x_1x_4x_5$ &$\Rightarrow$& 100110 & $\forall x_1x_4 \rightarrow x_5$ \\
$\exists x_1x_2x_6$ & $\Rightarrow$ & 110001 & $\forall x_1x_2 \rightarrow x_6$\\
$\exists x_3x_4x_5$ & $\Rightarrow$& 001110 & $\forall x_3x_4 \rightarrow x_5$\\
\\
\end{tabular}

We eliminate the last two tuples $\{110001, 001110\}$ as they are non-dominant. Therefore, A1 is
\begin{equation*}
\begin{array}{l}
111001\\
011110\\
110011\\
011011\\
100110\\
\end{array}
\end{equation*}
\newline
\noindent\textbf{[N1]~~}For each dominant distinguishing tuple of an existential expression, we construct the following four questions by replacing each distinguishing tuple with its children: 
\begin{equation*}
\begin{array}{cccc}
\exists x_1x_2x_3 (x_6) & \exists x_2x_3x_4 (x_5) & \exists x_1x_2x_5 (x_6) & \exists x_2x_3x_5x_6 \\
\hline
{\bf110001} & 111001       &  111001  & 111001\\
{\bf101001} & {\bf011010}  &  011110  & 011110\\
{\bf011001} & {\bf010110}  &{\bf110001} & 110011\\
011110 &      {\bf001110}  &{\bf100011} & \bf{011010}\\
110011 &      110011       &{\bf010011} & \bf{011001}\\
011011 &      011011       & 011011     & \bf{010011} \\
100110 &      100110       &  100110    & \bf{001011}\\
& & & 100110 \\
\end{array}
\end{equation*}
To avoid violating Horn expressions, we set affected head variables (in brackets) to true. We bold out the children of each dominant distinguishing tuple. Since the existential expression is not satisfied by any of the tuples in a question, the question is a non-answer.
\newline\newline
\noindent\textbf{[A2]~~}For each dominant universal Horn expression, we construct the following distinguishing tuples:

\begin{tabular}{lcl}
$\forall x_1x_4 \rightarrow x_5$ & $\Rightarrow$ & 100101 \\
$\forall x_3x_4 \rightarrow x_5$ & $\Rightarrow$ & 001101 \\
$\forall x_1x_2 \rightarrow x_6$ & $\Rightarrow$ & 110010 \\
\\
\end{tabular}

A2 questions consist of children of the universal distinguishing tuples:
\begin{equation*}
\begin{array}{ccc}
\forall x_1x_4 \rightarrow x_5 & \forall x_3x_4 \rightarrow x_5 & \forall x_1x_2 \rightarrow x_6 \\
\hline
111111 & 111111 & 111111 \\
100001 & 001001 & 100010 \\
000101 & 000101 & 010010 \\
\end{array}
\end{equation*}
\newline\newline
\noindent\textbf{[N2]~~}For each universal distinguishing tuple, we construct the following questions
\begin{equation*}
\begin{array}{ccc}
\forall x_1x_4 \rightarrow x_5 & \forall x_3x_4 \rightarrow x_5 & \forall x_1x_2 \rightarrow x_6 \\
\hline
111111 & 111111 & 111111 \\
100101 & 001101 & 110010 \\
\end{array}
\end{equation*}

\noindent\textbf{[A3]~~}First, we find existential expressions that dominate the guarantee clauses of a universal Horn expression.
$\exists x_2x_3x_4x_5$  dominates the guarantee clause $\exists x_3x_4x_5$ of $\forall x_3x_4 \rightarrow x_5$. Note that $\exists x_2x_3x_4x_5$ is implied by the expressions $\exists x_2x_3x_4 \ \forall x_3x_4 \rightarrow x_5$ in the query \ref{ax:exi-extend}. Second, we generate the question by generating search roots for the body $x_3x_4$ within the sub-lattice rooted at $011101$.
\begin{equation*}
\begin{array}{l}
111111\\
010101\\
111001\\
\end{array}
\end{equation*}
If in the intended query $x_5$ has another body in $x_2x_3x_4$ that is incomparable with $x_3x_4$ the above question will be a non-answer.
\newline\newline
\noindent\textbf{[A4]~~}The query has four non-head variables $\{x_1, x_2, x_3, x_4\}$. So we construct the following question.

\begin{equation*}
\begin{array}{l}
111111\\
011111\\
101111\\
110111\\
111011\\
\end{array}
\end{equation*}

\subsection{Completeness of a Verification Set}

\begin{figure}[htbp]
   \centering
   \includegraphics[width=01\linewidth]{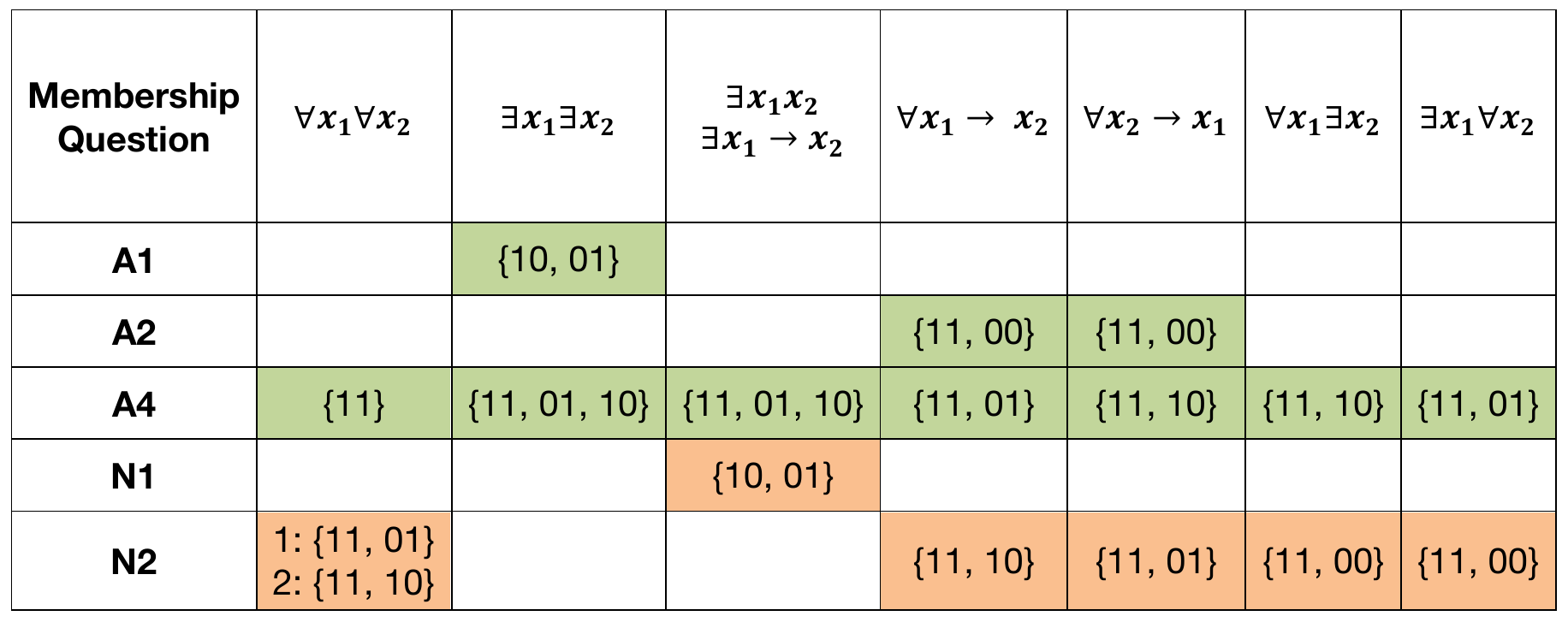}
   \caption{Verification sets for each role-preserving qhorn query on two variables}
   \label{fig:certificate}
\end{figure}
\begin{figure}[htbp]
   \centering
   \includegraphics[width=0.6\linewidth]{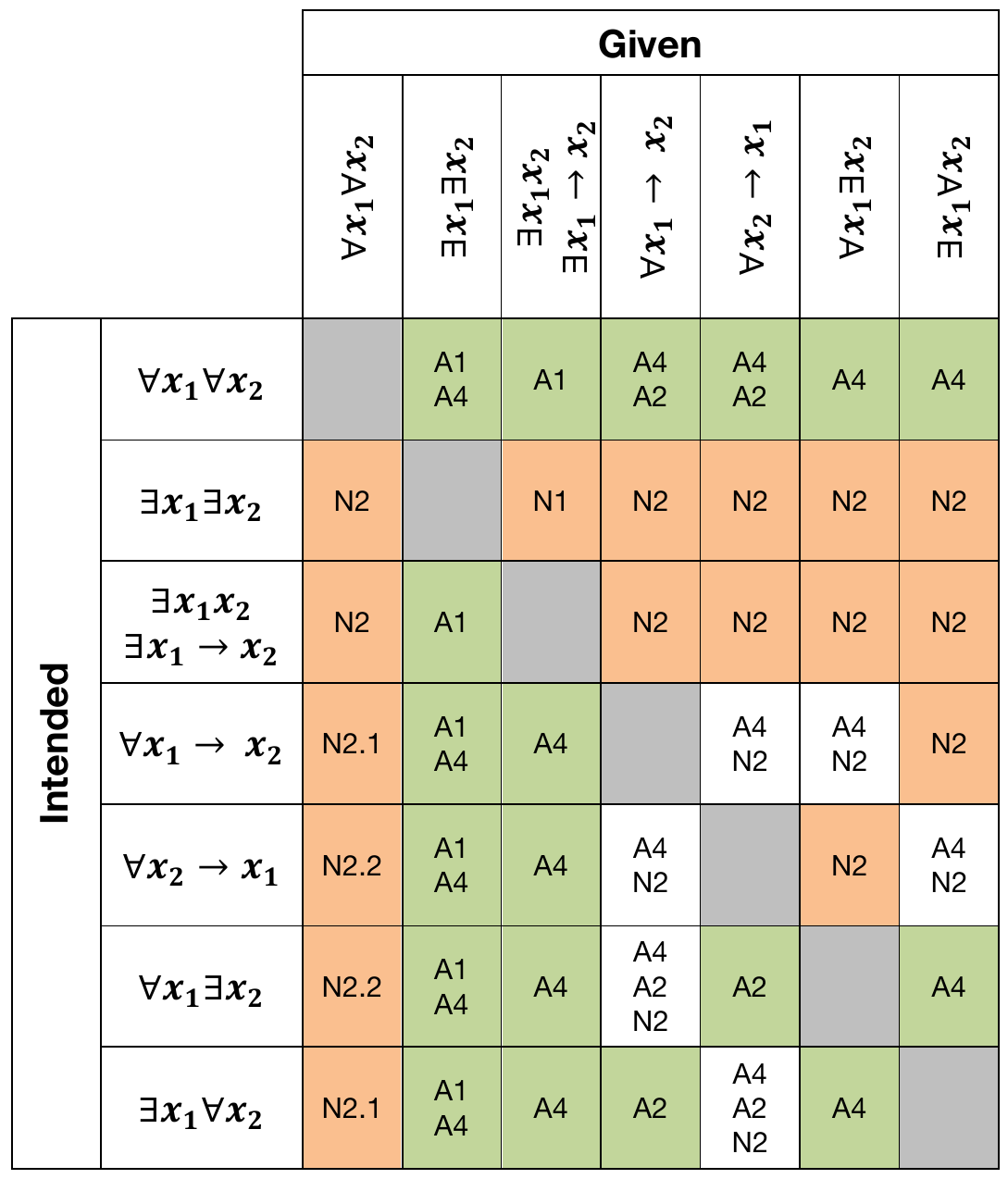}
   \caption{Membership questions that detect a difference between two specific role-preserving qhorn queries on two-variables.}
   \label{fig:matrix}
\end{figure}

\cref{fig:certificate} illustrates the verification sets of all role-preserving queries on two variables. \cref{fig:matrix} illustrates which membership questions in the verification sets enable the user to detect a discrepancy between the given query and the query they actually intended. Note that with only two variables, there is no need to generate A3 questions. These examples serve to demonstrate the completeness of the verification sets at least for role-preserving queries on two variables.

\begin{theorem}
\label{thm:vs}
A verification set with all membership questions of \cref{fig:two-table} surfaces semantic differences between the given query $q_g$ and the intended query $q_i$ by surfacing differences between the sets of distinguishing tuples of $q_g$ and $q_i$.
\end{theorem}

\emph{Proof:}
\emph{Case 1: $q_i$ and $q_g$ have different sets of dominant existential distinguishing tuples} then by Lemma \ref{lem:a1n1}, questions A1 and N1 surface differences in the sets of dominant existential distinguishing tuples of $q_g$ and $q_i$. 

\emph{Case 2: $q_i$ and $q_g$ have different sets of dominant universal distinguishing tuples} then
\begin{enumerate}[leftmargin=0.5cm,noitemsep]
\item Both $q_i$ and $q_g$ classify $h$ as a head variable. $q_i$ has a dominant universal Horn expression $C_i: \forall B_i \rightarrow h$ ($B$ is a set of body variables) and $q_g$ has dominant universal Horn expressions of the form $\forall B_g \rightarrow h$.
\begin{enumerate}[leftmargin=0.5cm,noitemsep,nolistsep]
\item If for \emph{any} $B_g$ in $q_g$, $B_i \subset B_g$ or $B_i \supset B_g$ then by Lemmas \ref{lem:a2} and \ref{lem:n2} questions A2 and N2 will surface this difference.
\item If for \emph{all} $B_g$ in $q_g$, $B_i$ and $B_g$ are incomparable then either (i) $C_i$'s guarantee clause dominates $q_g$'s existential expressions and $q_g$'s set of existential distinguishing tuples does not have the distinguishing tuple for $C_i$'s guarantee clause (See Case 1) or (ii) $C_i$'s guarantee clause is dominated by an existential expression in $q_g$ and by Lemma \ref{lem:a3} question A3 surfaces the difference.
\end{enumerate}
\item $h$ is a head variable in $q_i$ but is a non-head variable in $q_g$ then by Lemma \ref{lem:a4} question A4 surfaces the difference.\qed
\end{enumerate}

\begin{lemma}
\label{lem:a1n1}
Let $D_i$ be the set of $q_i$'s dominant existential distinguishing tuples and let $D_g$ be the set of $q_g$'s dominant existential distinguishing tuples; membership questions A1 and N1 surface $D_i \neq D_g$.
\end{lemma}

\emph{Proof:} An existential distinguishing tuple represents an inflection point: all questions constructed with tuples in the distinguishing tuple's upset are answers and all questions constructed with only tuples in the rest of the lattice are non-answers. We use this feature to detect if $D_i \neq D_g$. 

First, we define the following order relations over $D_i$ and $D_g$:
\begin{enumerate}[leftmargin=0.5cm,noitemsep]
\item $D_g \leq D_i$ if for every tuple $t_i \in D_i$, there exists a tuple $t_g \in D_g$ such that $t_g$ is in the upset of $t_i$.
\item $D_g \geq D_i$ if all tuples in $D_g$ are in the downset of $D_i$. 
\item $D_g || D_i$, otherwise, i.e. they are incomparable.
\end{enumerate}

Since $D_g \neq D_i$ only the following cases are possible:

\emph{Case 1: $D_g || D_i$ or $D_g > D_i$:} $D_g$ or membership question A1 is a non-answer to the user's intended query $q_i$. The user will detect the discrepancy as $D_g$ is presented as an answer in $q_g$'s verification set. 

\emph{Case 2: $D_g < D_i$}. Suppose all tuples in $D_g$ are in the upset of one of $D_i$'s tuples. Let $D_g(t)$ be the set of distinguishing tuples where we replace $t \in D_g$ with its children. There are $|D_g| = O(k)$ such sets. These sets form membership questions N1. For any $t \in D_g$, $D_g(t)$ is always a non-answer to $q_g$. However, for at least one tuple $t$, $D_g(t)$ is an answer to $q_i$. This is because if $D_g < D_i$ then at least one of $D_i$'s tuples is a descendant of one of $D_g$'s tuples, in which case $D_g(t)$ is still in the upset of that tuple and thus an answer. The user will detect the discrepancy as $D_g(t)$ is presented as a non-answer in $q_g$'s verification set. \qed

Like existential distinguishing tuples, universal distinguishing tuples represent an inflection point. All tuples in the upset of the universal distinguishing tuple are non-answers (as all of $h$'s body variables are true but $h$ is false). All descendants of the universal distinguishing tuple are answers (as no complete set of $h$'s body variables is true). 

Let $t_i$ be $q_i$'s universal distinguishing tuple for an expression on the head variable $h$. Let $t_g$ be one of $q_g$'s universal distinguishing tuples for expressions on the head variable $h$.
We define the following order relations between $t_i$ and $t_g$:
\begin{enumerate}[leftmargin=0.5cm,noitemsep] 
\item $t_i \leq t_g$ if $t_i$ is in the upset of $t_g$. 
\item $t_i \geq t_g$ if $t_i$ is in the downset of $t_g$. 
\item $t_i || t_g$ if $t_i$ and $t_g$ are incomparable.
\end{enumerate}
Consider two distinct (dominant) tuples $t_{g_1}$ and $t_{g_2}$ of the given query. By qhorn's equivalence rules\cref{sec:axioms} queries $t_{g_1}$ and $t_{g_2}$ are incomparable ($t_{g_1} || t_{g_2}$). Consequently, for any two distinct tuples both $t_i < t_{g_1}$ and $t_i > t_{g_2}$ cannot hold.

\begin{lemma}
\label{lem:a2}
Membership question A2 detects $t_i > t_g$. 
\end{lemma}

\emph{Proof:}
Suppose, $q_g$ has one universal distinguishing tuple $t_g$ such that $t_i > t_g$. Then the membership question A2 that consists of the all-true tuple and $t_g$'s children is an answer for $q_g$ as none of $t_g$'s children have all the body variables set to true, so the head variable can be false. If $t_i > t_g$ then $q_i$'s universal Horn expression on $h$ has a strict subset of the body variables represented by $t_g$. Therefore, in at least one of $t_g$'s children, all of $t_i$'s body variables are set to true and $h$ is still false. Thus, A2 is a non-answer to $q_i$. For all other universal distinguishing tuples $t_g$ of $q_g$, either $t_i > t_g$ or $t_i || t_g$. If $t_i || t_g$ then A2 is still an answer. \qed

\begin{lemma}
\label{lem:n2}
Membership question N2 detects $t_i < t_g$. 
\end{lemma}

\emph{Proof:}
Suppose, $q_g$ has one universal distinguishing tuple $t_g$ such that $t_i < t_g$. Then the membership question N2 that consists of the all-true tuple and $t_g$ is a non-answer for $q_g$ as $t_g$ has all body variables set to true but the head variable $h$ is false. If $t_i < t_g$ then $q_i$'s universal Horn expression on $h$ has a strict superset of the body variables represented by $t_g$. Therefore, $t_g$ does not have all body variables set to true and $h$ can be false. Thus, N2 is an answer to $q_i$.

For all other universal distinguishing tuples $t_g$ of $q_g$, either $t_i < t_g$ or $t_i || t_g$. If $t_i || t_g$ then N2 is still a non-answer. \qed

\begin{lemma}
\label{lem:a3}
If
\begin{itemize}[leftmargin=0.5cm,noitemsep,nolistsep]
\item $h$ is a head variable in $q_i$ and $q_g$. 
\item $q_i$ has a dominant universal Horn expression $\forall M \rightarrow h$ which $q_g$ does not have.
\item $q_g$ has universal Horn expressions $\forall B_1 \rightarrow h \ ... \forall B_\theta \rightarrow h$.
\item $B_i || M$ for $i = 1 ... \theta$
\item $q_g$ has an existential expression on $C$ variables ($\exists \ C$) such that $C \supseteq M$ and $C \supset B_i$ for $i = 1...\theta$
\end{itemize}
then A3 surfaces a missing universal Horn expression ($\forall M \rightarrow h$) from $q_g$.
\end{lemma}

\emph{Proof:} Consider $q_g$'s universal Horn expressions whose guarantee clauses are dominated by $\exists \ C$: 
\[\forall B_1 \rightarrow h, \forall B_2 \rightarrow h, ... \forall B_\theta \rightarrow h\] 
such that $B_i \subset C$ for $i = 1 ... \theta$. 
To build A3, we set one body variable from each of $B_1, ..., B_\theta$ to false, the remaining variables in $C$ to true and $h$ to false. There are $|B_1| \times |B_2| \times ... \times |B_\theta| = O(n^\theta)$ such tuples. A3 now consists of all such tuples and the all-true tuple.

A3 acts like the search phase of the learning algorithm that looks for new universal Horn expressions\cref{sec:new-horn}. A3 is a non-answer for $q_i$ as at least one of the tuples has all variables in $M$ set to true (because $M || B_i$ for $i = 1 ... \theta$) and $h$ to false, thus violating $\forall M \rightarrow h$. \qed

\begin{lemma}
\label{lem:a4}
If $h$ is a head variable in $q_i$ but not in $q_g$ then question A4 surfaces the difference.
\end{lemma}

\emph{Proof:}
The all-true tuple satisfies all existential expressions in $q_g$. For each body variable $x$ in $q_g$, A4 has a tuple where $x$ is false and all other variables are true. If $x$ is a head variable in $q_i$, then A4 should be a non-answer. \qed

This concludes the proof of \cref{thm:vs}

\section{Related Work}

\textbf{Learning \& Verifying Boolean Formula: } 
Our work is influenced by the field of computational learning theory. Using membership questions to learn Boolean formulas was introduced in 1988~\cite{angluin-concept}. Angluin et al. demonstrated the polynomial learnability of conjunctions of (non-quantified) Horn clauses using membership questions and a more powerful class of questions known as equivalence questions~\cite{angluin-horn}. The learning algorithm runs in time $O(k^2n^2)$ where $n$ is the number of variables and $k$ is the number of clauses. Interestingly, Angluin proved that there is no PTIME algorithm for learning conjunctions of Horn clauses that only uses membership questions. Angluin et al.'s algorithm for learning conjunctions of Horn formula was extended to learn first-order Horn expressions~\cite{khardon, haussler}. First-order Horn expressions contain quantifiers. We differ from this prior work in that in qhorn we quantify over tuples of an object's nested relation; we do not quantify over the values of variables. Our syntactic restrictions on qhorn have counterparts in Boolean formulas. Both qhorn-1 and \emph{read-once} Boolean formulas~\cite{angluin-readonce} allow variables to occur at most once. Both role-preserving qhorn queries and \emph{depth-1 acyclic Horn formulas}~\cite{sloan} do not allow variables to be both head and body variables.

Verification sets are analogous to the \emph{teaching sequences} of Goldman and Kearns~\cite{goldman}. A teaching sequence is the smallest sequence of classified examples a teacher must reveal to a learner to help it uniquely identify a target concept from a concept class. Prior work provides algorithms to determine the teaching sequences for several classes of Boolean formula~\cite{specifnum, goldman, shinohara} but not for our class of qhorn queries.

\textbf{Learning in the Database Domain: }
Two recent works on example-driven database query learning techniques --- Query by Output (QBO)~\cite{qbo} and Synthesizing View Definitions (SVD)~\cite{svd} --- focus on the problem of learning a query $Q$ from a given input database $D$, and an output view $V$. There are several key differences between this body of work and ours. First, QBO and SVD perform as decision trees; they infer a query's propositions so as to split $D$ into tuples in $V$ and tuples not in $V$. We assume that users can provide with us the propositions, so we focus on learning the structure of the query instead. Second, we work on a different subset of queries: QBO infers select-project-join queries and SVD infers unions of conjunctive queries. Learning unions of conjunctive queries is equivalent to learning $k$-term Disjunctive Normal Form (DNF) Boolean formulae~\cite{kearns}. We learn conjunctions of \emph{quantified} Horn formulae. Since our target queries operate over objects with nested-sets of tuples instead of flat tuples, we learn queries in an exponentially larger query and data space. Finally, QBO and SVD work with a complete mapping from input tuples to output tuples. Our goal, however, is to learn queries from the smallest possible mapping of input to output objects, as it is generally impractical for users to label an entire database of objects as answers or non-answers. We point out that we synthesize our input when constructing membership questions, thus we can learn queries independent of the peculiarities of a particular input database $D$. 

Using membership (and more powerful) questions to learn concepts within the database domain is not novel. For example,
Cate, Dalmau and Kolaitis use membership and equivalence questions to learn schema mappings~\cite{schemalearn}. A schema mapping is a collection of first-order statements that specify the relationship between the attributes of a source and a target schema. Another example is Staworko's and Wieczorek's work on using example XML documents given by the user to infer XML queries~\cite{xmllearn}. In both these works, the concept class learned is quite different from the qhorn query class.

\textbf{The Efficacy of Membership Questions}
Learning with membership questions is also known as \emph{Active Learning}. Active learning elicits several criticisms due to mixed or negative results in some learning problems. We wish to address two of the main criticisms:

\begin{enumerate}[leftmargin=0.5cm]
\item \emph{Arbitrary Examples.} Early work by Lang and Baum~\cite{lang} used membership questions to train a neural network to recognize hand-written digits. They discovered that users couldn't reliably respond to the questions --- images of artificially synthesized hybrids of two digits\footnote{Later work by Kudo et al. demonstrates how appropriately constructed membership questions can boost the performance of character recognition algorithms~\cite{kudo}.}. 
This canonical negative result does not apply to our work. We synthesize examples from the actual data domain. Moreover, if we have a rich database, we can select instances from the database that match our synthesized Boolean tuples instead of synthesizing the data tuples. 
\item \emph{Noisy Users.} The criticism made here is that (i) users may not know have a clear idea of what constitutes a positive (answer) or negative (non-answer) example or (ii) users make mistakes. 
In query specification tasks, users typically have a clear idea of what they are looking for. This contrasts with data exploration tasks, where users search the database without a well-defined selection criteria. A good user-interface can ameliorate the second issue. For example, if we provide users with a history of all their responses to the different membership questions, users can double-check their responses and change an incorrect response. This triggers the query learning algorithm to restart query learning from the point of error.
\end{enumerate}

A survey by Settles discusses recent advances and challenges in active learning~\cite{settles}.

\section{Conclusion \& Future Work}

In this paper, we have studied the learnability of a special class of Boolean database queries --- qhorn. We believe that other quantified-query classes (other than conjunctions of quantified Horn expressions) may exhibit different learnability properties. Mapping out the properties of different query classes will help us better understand the limits of example-driven querying. 
In our learning/verification model, we made the following assumptions: (i) the user's intended query is either in qhorn-1 or role-preserving qhorn, (ii) the data has at most one level nesting. We plan to design algorithms to verify that the user's query is indeed in qhorn-1 or role-preserving qhorn. 
We have yet to analyze the complexity of learning queries over data with multiple-levels of nesting. In such queries, a single expression can have several quantifiers.

We plan to investigate \emph{Probably Approximately Correct} learning: we use randomly-generated membership questions to learn a query with a certain probability of error~\cite{valiant}. We note that membership questions provide only one bit of information --- a response to membership question is either `answer' (1) or `non-answer' (0). We plan to examine the plausibility of constructing other types of questions that provide more information bits but still maintain interface usability. One possibility is to ask questions to directly determine how propositions interact\footnote{We thank our anonymous reviewer for this suggestion.} such as: ``do you think $p_1$ and $p_2$ both have to be satisfied by at least one tuple?" or ``when does $p_1$ have to be satisfied?''

Finally, we see an opportunity to create efficient \emph{query revision} algorithms. Given a query which is \emph{close} to the user's intended query, our goal is to determine the intended query through few membership questions --- polynomial in the distance between the given query and the intended query. Efficient revision algorithms exist for (non-quantified) role-preserving Horn formula~\cite{sloan}. The Boolean-lattice provides us with a natural way to measure how close two queries are: the distance between the distinguishing tuples of the given and intended queries. 
\newline\newline
\noindent\textbf{Acknowledgments} Partial funding provided by NSF Grants CCF-0963922, CCF-0916389, CC-0964033 and a Google University Research Award.

\balance 
\bibliographystyle{abbrv}
\bibliography{pods13_compact}

\end{document}